\documentclass[aps, pra, twocolumn, superscriptaddress]{revtex4-2}

\usepackage[]{amsmath}
\usepackage{amssymb}
\usepackage{mathrsfs}

\usepackage{fontawesome}

\usepackage[]{graphicx}
\graphicspath{{figures/}}
\usepackage[]{mathtools}
\usepackage[]{bm}
\usepackage[]{braket}
\usepackage{esint}
\usepackage{ulem}
\usepackage{cancel}
\usepackage[caption=false]{subfig}

\newcommand{\figref}[1]{\figurename~\ref{#1}}

\newcommand{\gl}{{\raisebox{.15em}[0em][0em]{$\scriptscriptstyle{>}$}\hspace{-.52em}\raisebox{-.15em}[0em][0em]{$\scriptscriptstyle{<}$}}}
\newcommand{\ssl}{{\scriptscriptstyle{<}}}
\newcommand{\ssg}{{\scriptscriptstyle{>}}}
\newcommand{\ssm}{{\scriptscriptstyle{-}}}
\newcommand{\ssp}{{\scriptscriptstyle{+}}}
\newcommand{\Inc}{\ssm\ssg}
\newcommand{\Refl}{\ssp\ssg}
\newcommand{\Tra}{\ssm\ssl}
\newcommand{\cc}{\raisebox{.3em}{$\scriptstyle{*}$}}

\newcommand{\inv}{\raisebox{.3em}{$\scriptscriptstyle{-1}$}}
\newcommand{\sgn}{\operatorname{sgn}}

\DeclareMathOperator*{\SumInt}{%
  \mathchoice%
  {\ooalign{$\displaystyle\sum$\cr\hidewidth$\displaystyle\int$\hidewidth\cr}}
  {\ooalign{\raisebox{.14\height}{\scalebox{.7}{$\textstyle\sum$}}\cr\hidewidth$\textstyle\int$\hidewidth\cr}}
  {\ooalign{\raisebox{.2\height}{\scalebox{.6}{$\scriptstyle\sum$}}\cr$\scriptstyle\int$\cr}}
  {\ooalign{\raisebox{.2\height}{\scalebox{.6}{$\scriptstyle\sum$}}\cr$\scriptstyle\int$\cr}}
}

\usepackage{cases}
\usepackage[]{comment}

\begin{document}

\title{
  Calculating spatiotemporally modulated surfaces:
  a dynamical differential formalism
}

\author{Daigo Oue}
\affiliation{The Blackett Laboratory, Department of Physics, Imperial College London, Prince Consort Road, Kensington, London SW7 2AZ, United Kingdom}
\author{Kun Ding}
\affiliation{The Blackett Laboratory, Department of Physics, Imperial College London, Prince Consort Road, Kensington, London SW7 2AZ, United Kingdom}
\affiliation{Department of Physics, State Key Laboratory of Surface Physics, and Key Laboratory of Micro and Nano Photonic Structures (Ministry of Education), Fudan University, Shanghai 200438, China}
\author{J. B. Pendry}
\affiliation{The Blackett Laboratory, Department of Physics, Imperial College London, Prince Consort Road, Kensington, London SW7 2AZ, United Kingdom}

\date{\today}

\begin{abstract}
  Electromagnetic waves in a system with a space and time dependent boundary experience both diffraction and Doppler-like frequency conversion.
  In order to analyse such situations,
  conventional methods call for either the eigenmodes or the dyadic Green's function in space and time dependent media.
  Here, 
  we propose a dynamical differential method which does not require either of them.
  Our method utilises a dynamical coordinate transformation in order to simplify the calculation of the optical response of the space and time dependent system.
  We reveal that the diffraction symmetry is broken in the presence of traveling-wave type spatiotemporal modulation.
\end{abstract}

\maketitle

\section{Introduction}
\label{sec:introduction}
Recently,
spatiotemporal modulation of bulky media has been studied both theoretically and experimentally
\cite{%
  sounas2017non,
  caloz2018electromagnetic,
  shaltout2019spatiotemporal,
  galiffi2019broadband,
  huidobro2019fresnel%
}.
The modulation breaks reciprocity in the media and brings about unique phenomena such as light amplification and the Fresnel drag.
Keeping these in mind, 
in this work,
we analyse the light scattering by spatiotemporally modulated surfaces.

One way to calculate the light scattering is to use the dyadic Green's function in the system in question.
Once we obtain the dyadic Green's function in the system,
we can perturbatively calculate the scattered field from incident field and thus the scattering matrix,
\cite{%
  goedecke1988scattering,
  Yurkin_2007%
}.

Another approach is the boundary matching method.
In this approach,
the incident and scattered fields are expanded in series of the eigenmodes in the system,
and the field continuity conditions are imposed at the boundaries,
which lead to simultaneous equations.
The inversion of the equation system yields the scattering matrix.
Calculation of the Mie coefficients \cite{mie1908beitrage} is one of the popular examples where boundary matching approach is taken.

Here, we consider the boundary matching at a dynamically deformed boundary with the help of a differential formalism.
The differential formalism is originally proposed by Chandezon et al. to calculate dielectric gratings
\cite{%
  chandezon1980new,
  chandezon1982multicoated,
  li1994multilayer%
}.
Since this method allows to directly match the boundary conditions at structured interfaces,
the geometry is properly captured.
One significance of the differential formalism is its convergence.
When the formalism is numerically implemented,
it shows steady convergence even in the presence of sharp edges
\cite{%
  li1996improvement,
  li1996use%
}.
This is why it has been used to calculate not only dielectric gratings but also dispersive, lossy, anisotropic gratings
\cite{%
  barnes1995photonic,
  harris1995differential,
  barnes1996physical,
  harris1996conical,
  kitamura2013hermitian,
  murtaza2017study%
}.
In the current work,
we extend the formalism to dynamical systems so that we can take the motion of the interface as well as the geometry.

There is a related approach based on conformal transformations proposed by Ward and Pendry
\cite{ward1996refraction},
in which a surface structure is encoded into a conformal mappings.
This approach is also widely used in optics and plasmonics communities
\cite{%
  leonhardt2006optical,
  liu2010transformational,
  vakil2011transformation,
  xu2015conformal,
  pendry2015transforming,
  pendry2019computing%
}.
Compared to the conformal approach,
the method utilised in this paper is more straightforward when it comes to analysing systems with time dependent boundaries because we can directly use time-dependent surface profile instead of finding corresponding conformal transformation.

This paper is organised as following.
In Sec. \ref{sec:differential},
we extend a differential formalism so as to enable the electromagnetic field analysis in the presence of a dynamically modulated boundary.
In Sec. \ref{sec:numerial},
we numerically implement the dynamical differential formalism.
We will also see the calculation is consistent with effective medium description.
The conclusion is drawn in Sec. \ref{sec:conclusion}.

\section{Dynamical differential formalism}
\label{sec:differential}
\begin{figure}[htbp]
  \centering
  \includegraphics[width=\linewidth]{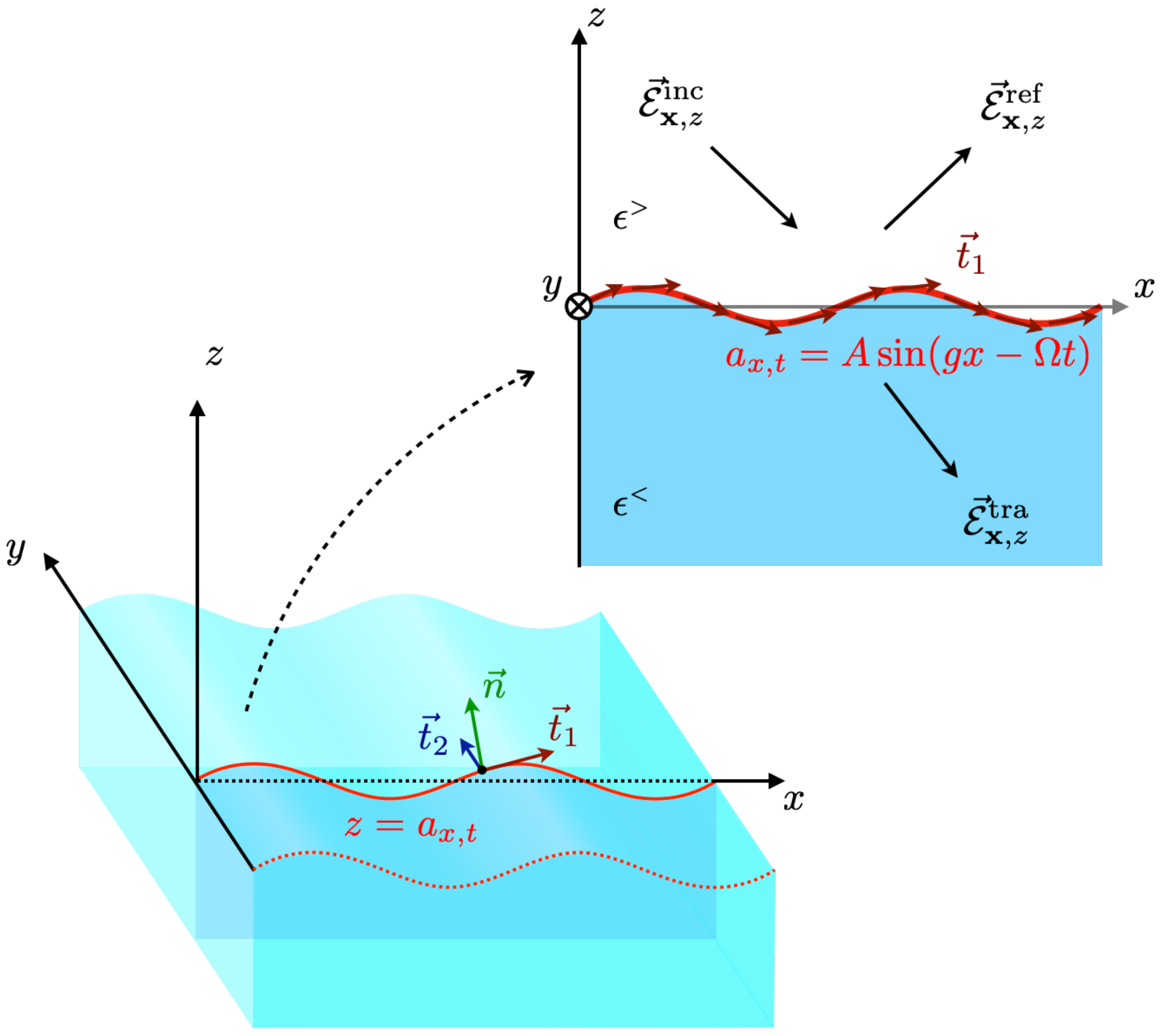}
  \caption{
    Spatiotemporally modulated surface.
    The surface is weakly modulated in space and time.
    The corrugation function is sinusoidal and given by Eq. \eqref{eq:boundary}.
    We consider in-plane incidence of electromagnetic field and analyse the reflection and the transmission at the surface.
    The red and blue arrows $\vec{t}_{1,2}$ are the tangential vectors of the surface at each point.
    The permittivities of medium and lower media are $\epsilon^{\gl}$, 
    respectively.
    The permeability is assumed to be unity in both media, $\mu^{\gl} = 1$.
  }
  \label{fig:sketch}
\end{figure}

As a simple example of the dynamically modulated surfaces,
we consider a sinusoidally modulated boundary shown in \figref{fig:sketch}.

The profile of the boundary is given by
\begin{align}
  a_\mathbf{x} 
  = 
  A \sin(\mathbf{q}\cdot\mathbf{x})
  = A \sin \left[g\left(x - \frac{\Omega}{g} t\right)\right]
  \label{eq:boundary}
\end{align}
Here, we have introduced three-component vectors $\mathbf{q} = \{ g, 0, -i\Omega/c\}$ and $\mathbf{x} = \{x, y, -ict\}$,
where $A$, $g$ and $\Omega$ are the strength of the modulation, the spatial and temporal frequencies of the modulation, respectively.
The surface is not physically moving in the $x$ direction,
but its profile is shifting at the phase velocity $\Omega/g$ that can exceed the speed of light.

The permittivity distribution is given by means of the boundary profile,
\begin{align}
  \epsilon_{\mathbf{x},z}
  &= 
  \epsilon^\ssl
  \Theta(a_\mathbf{x}-z) 
  +
  \epsilon^\ssg
  \Theta(z-a_\mathbf{x})
  \\
  &= 
  \alpha
  \Theta(a_\mathbf{x}-z)
  +
  \epsilon^\ssg,
\end{align}
where
$\Theta$ represents the Heaviside unit step function,
and 
$\alpha \equiv (\epsilon^\ssl-\epsilon^\ssg)$
corresponds to the permittivity difference.

We can calculate tangential and normal vectors of the surface,
$\vec{t}_{1,2}$ and $\vec{n}$, 
by taking the partial derivatives of the boundary profile \eqref{eq:boundary},
\begin{align}
  \begin{cases}{}
    \vec{t}_{1}
    =
    \cfrac{\vec{u}_x + a_\mathbf{x}' \vec{u}_z}{\sqrt{1+{a_\mathbf{x}'}^2}},
    \quad
    \vec{t}_2
    =
    \vec{u}_y.
    \\
    \vec{n} 
    = 
    \vec{t}_1 \times \vec{t}_2
    =
    \cfrac{-a_\mathbf{x}' \vec{u}_x + \vec{u}_z}{\sqrt{1+{a_\mathbf{x}'}^2}},
  \end{cases}
  \label{eq:t_1,2,n}
\end{align}
where $a_\mathbf{x}' = (\partial/\partial x)a_\mathbf{x}$ is the partial derivative of the boundary profile in the $x$ direction,
and $\vec{u}_{x,y,z}$ are the unit vectors in the $x, y$ and $z$ directions.
We use those tangential and normal vectors when considering the field continuity conditions or surface integrations over the modulated interface.

\subsection{Fourier expansion of fields}
\label{subsec:Fourier}
Since the upper and lower spaces of the boundary are homogeneous dielectrics,
we can unambiguously expand the incident, reflected and transmitted fields in a series of plane waves,
\begin{align}
  \begin{cases}{}
    \vec{\mathcal{E}}{}_{\mathbf{x},z}^\mathrm{inc} 
    = \displaystyle{\int_{\mathbf{k}}}
    e^{i\mathbf{k}\cdot \mathbf{x}}
    \vec{E}{}_{\mathbf{k},z}^{\Inc},
    &
    \vec{\mathcal{H}}{}_{\mathbf{x},z}^\mathrm{inc} 
    = \displaystyle{\int_{\mathbf{k}}}
    e^{i\mathbf{k}\cdot \mathbf{x}}
    \vec{H}{}_{\mathbf{k},z}^{\Inc},
    \vspace{.5em}
    \\
    \vec{\mathcal{E}}{}_{\mathbf{x},z}^\mathrm{ref} 
    = \displaystyle{\int_{\mathbf{k}}}
    e^{i\mathbf{k}\cdot \mathbf{x}}
    \vec{E}{}_{\mathbf{k},z}^{\Refl},
    &
    \vec{\mathcal{H}}{}_{\mathbf{x},z}^\mathrm{ref} 
    = \displaystyle{\int_{\mathbf{k}}}
    e^{i\mathbf{k}\cdot \mathbf{x}}
    \vec{H}{}_{\mathbf{k},z}^{\Refl},
    \vspace{.5em}
    \\
    \vec{\mathcal{E}}{}_{\mathbf{x},z}^\mathrm{tra} 
    = \displaystyle{\int_{\mathbf{k}}}
    e^{i\mathbf{k}\cdot \mathbf{x}}
    \vec{E}{}_{\mathbf{k},z}^{\Tra},
    &
    \vec{\mathcal{H}}{}_{\mathbf{x},z}^\mathrm{tra} 
    = \displaystyle{\int_{\mathbf{k}}}
    e^{i\mathbf{k}\cdot \mathbf{x}}
    \vec{H}{}_{\mathbf{k},z}^{\Tra}.
    \vspace{.5em}
  \end{cases}
  \label{eq:E,H_lnc,ref,tra}
\end{align}
The superscripts on the left hand side
($
\Lambda 
= 
\mathrm{inc},\ 
\mathrm{ref},\ 
\mathrm{tra}
$)
are labels which identify modes in the real space and the time domain.
The two superscripts on the right hand side
($
\sigma = \pm,\ 
\tau = \gl
$)
are corresponding labels in the reciprocal space.
While $\sigma$ specifies in which direction the mode propagates,
$\tau$ specifies in which medium the mode lives.
The fields in the real space,
$\vec{\mathcal{E}}_{\mathbf{x},z}^{\Lambda}$ 
and
$\vec{\mathcal{H}}_{\mathbf{x},z}^{\Lambda}$,
are given by means of complex-valued Fourier components,
$\vec{E}{}_{\mathbf{k},z}^{\sigma\tau}$ 
and 
$\vec{H}{}_{\mathbf{k},z}^{\sigma\tau}$.
Note that we have introduced a reciprocal vector 
$\mathbf{k} = \{k_x, k_y, -ik_0 = -i\omega/c\}$.
Each Fourier component satisfies
\begin{align}
  \vec{E}{}_{\mathbf{k},z}^{\sigma\tau}
  &= 
  \vec{E}{}_{-\mathbf{k},z}^{\sigma\tau\cc},
  \quad
  \vec{H}{}_{\mathbf{k},z}^{\sigma\tau}
  = 
  \vec{H}{}_{-\mathbf{k},z}^{\sigma\tau\cc}.
\end{align}
These conditions are required because 
$\vec{\mathcal{E}}_{\mathbf{x},z}^{{\sigma\tau}}$
and 
$\vec{\mathcal{H}}_{\mathbf{x},z}^{{\sigma\tau}}$,
are real-valued.
We employ a shorthand notation for an integral operation,
\begin{align}
  \int_{\mathbf{k}}
  =
  \iint_{-\infty}^{+\infty} \frac{\mathrm{d}k_x \mathrm{d}k_y}{(2\pi)^2}
  \int_{-\infty}^{+\infty} \frac{\mathrm{d}\omega}{2\pi}.
\end{align}

We substitute Eq. \eqref{eq:E,H_lnc,ref,tra} into Maxwell's equations to get the Helmholtz equation,
\begin{align}
  \left[
    -\frac{\partial^2}{\partial z^2} 
    - 
    \left( 
      \epsilon^\tau 
      {k_0}^2
      -
      {k_\parallel}^2 
    \right)
  \right]
  \vec{E}{}_{\mathbf{k},z}^{\sigma\tau} = 0,
\end{align}
where we have defined 
$k_\parallel = \sqrt{{k_x}^2+{k_y}^2}$.
This differential equation can be easily solved,
\begin{align}
  \vec{E}{}_{\mathbf{k},z}^{\sigma\tau}
  &= 
  e^{i\sigma K_{\mathbf{k}}^\tau z}
  \vec{E}{}_{\mathbf{k},0}^{\sigma\tau}.
  \label{eq:sol}
\end{align}
Here, 
$\vec{E}{}_{\mathbf{k},0}^{\sigma\tau}$ is a complex-valued vector at $z=0$,
which determines the amplitude and the polarisation,
and we define the wavenumber in the $z$ direction,
\begin{align}
  K_{\mathbf{k}}^\tau 
  &= 
  \sgn(\omega) \operatorname{Re}
  \sqrt{
    \epsilon^\tau 
    {k_0}^2
    - {k_\parallel}^2
  }
  +
  i\operatorname{Im}
  \sqrt{
    \epsilon^\tau
    {k_0}^2
    -
    {k_\parallel}^2 
  }.
  \label{eq:K_alpha^tau}
\end{align}
We have added the prefactor $\sgn(\omega)$ at the real part in order to take into account that the propagation direction of a wave reverses when the sign of the frequency is flipped \cite{pendry2008time}.
Remind that the wavenumber satisfies
$
K_{-\mathbf{k}}^{\tau\cc}
= -K_{\mathbf{k}}^\tau
$.

In our dielectric medium,
the field is divergenceless 
(i.e. $\nabla \cdot \vec{\mathcal{E}}{}_{\mathbf{x},z}^{\Lambda}=0$)
so that we have a transversality condition,
\begin{align}
  i(
  k_x \vec{u}_x 
  + k_y \vec{u}_y 
  + \sigma K_{\mathbf{k}}^\tau \vec{u}_z
  )
  \cdot 
  \vec{E}{}_{\mathbf{k},0}^{\sigma\tau} = 0.
\end{align}
We can choose two orthonormal basis vectors satisfying this condition,
\begin{align}
  \vec{e}_{\lambda,\mathbf{k}}^{\hspace{.2em}\sigma\tau}
  &=
  \begin{cases}{}
    \cfrac{\operatorname{sgn}(\omega)\vec{k} \times \vec{u}_z}{|\vec{k} \times \vec{u}_z|}
    &
    (\lambda = s),
    \\
    \cfrac{
      \operatorname{sgn}(\omega)\vec{k} 
      \times 
      \operatorname{sgn}(\omega)\vec{k}
    \times \vec{u}_z
  }{|\vec{k} \times \vec{k} \times \vec{u}_z|}
    &
    (\lambda = p),
  \end{cases}
  \label{eq:e_lambda}
\end{align}
where $\vec{k} = k_x \vec{u}_x + k_y \vec{u}_y + \sigma K_{\mathbf{k}}^\tau \vec{u}_z$ is the wavevector of a mode labeled by
$\sigma$, 
$\tau$
and 
$\mathbf{k} = \{k_x,k_y,-ik_0\}$.

The complex-valued vector is expressed by the linear combination of the basis vectors,
\begin{align}
  \vec{E}{}_{\mathbf{k},0}^{\sigma\tau}
  &= 
  \sum_{\lambda=s,p} 
  E_{\lambda,\mathbf{k}}^{\sigma\tau}
  \vec{e}_{\lambda,\mathbf{k}}^{\hspace{.2em}\sigma\tau}.
\end{align}
Since the basis vectors satisfies
$
\vec{e}_{\lambda,-\mathbf{k}}^{\hspace{.2em}\sigma\tau\cc}
= \vec{e}_{\lambda,\mathbf{k}}^{\hspace{.2em}\sigma\tau}
$,
the constraint on the scalar coefficient is
\begin{align}
  E_{\lambda,\mathbf{k}}^{\sigma\tau}
  = 
  E_{\lambda,-\mathbf{k}}^{\sigma\tau\cc}.
  \label{eq:E_RFC}
\end{align}

We can perform a similar calculation for the magnetic field.
Let us summarise the Fourier components of electric and magnetic fields,
\begin{align}
  \begin{cases}{}
    \vec{E}{}_{\mathbf{k},z}^{\sigma\tau}
    =
    e^{i\sigma K_{\mathbf{k}}^\tau z}
    \vec{E}{}_{\mathbf{k},0}^{\sigma \tau},
  &
  \vec{E}{}_{\mathbf{k},0}^{\sigma \tau}
  =
  \displaystyle{\sum_\lambda}
  E_{\lambda,\mathbf{k}}^{\sigma\tau}
  \vec{e}_{\lambda,\mathbf{k}}^{\hspace{.2em}\sigma\tau},
  \vspace{.5em}
  \\
  \vec{H}{}_{\mathbf{k},z}^{\sigma\tau}
  =
  e^{i\sigma K_{\mathbf{k}}^\tau z}
  \vec{H}{}_{\mathbf{k},0}^{\sigma \tau},
  &
  \vec{H}{}_{\mathbf{k},0}^{\sigma \tau}
  =
  \displaystyle{\sum_\lambda}
  H_{\lambda,\mathbf{k}}^{\sigma\tau}
  \vec{h}_{\lambda,\mathbf{k}}^{\hspace{.2em}\sigma\tau},
  \end{cases}
  \label{eq:E,H_z,alpha^sigmatau}
\end{align}
where we have introduced another basis for the magnetic field for convenience,
\begin{align}
  \begin{pmatrix}
    \vec{h}_{s,\mathbf{k}}^{\sigma\tau}
    \\
    \vec{h}_{p,\mathbf{k}}^{\sigma\tau}
  \end{pmatrix}
  =
  \begin{pmatrix}
    0 & 1
    \\
    -1 & 0
  \end{pmatrix}
  \begin{pmatrix}
    \vec{e}_{s,\mathbf{k}}^{\hspace{.2em}\sigma\tau}
    \\
    \vec{e}_{p,\mathbf{k}}^{\hspace{.2em}\sigma\tau}
  \end{pmatrix}.
\end{align}

\subsection{Boundary matching equations}
Let us derive boundary matching equations,
which are used to calculate the reflection and transmission matrices.
\begin{figure}[htbp]
  \centering
  \includegraphics[width=.7\linewidth]{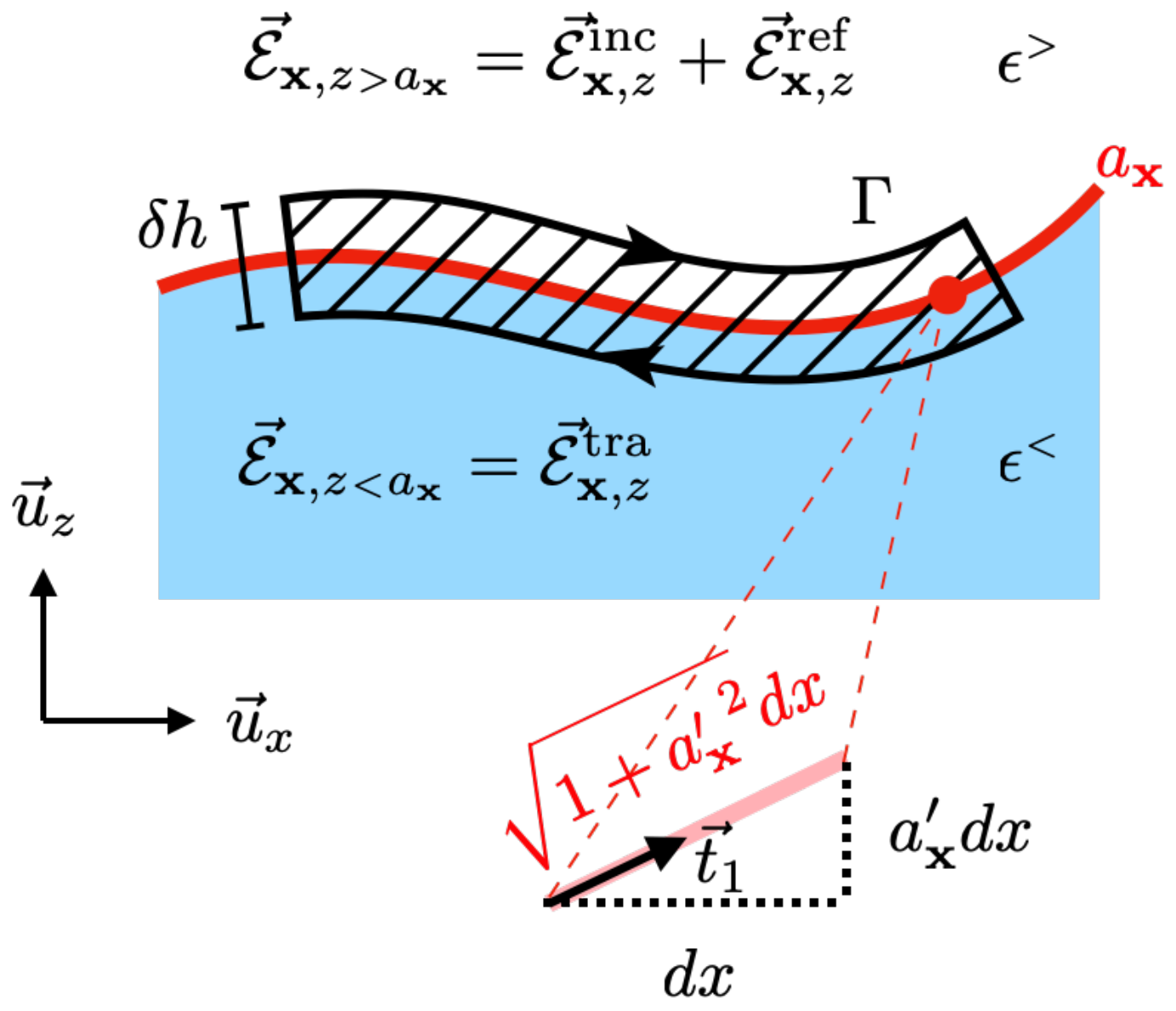}
  \caption{
    Path of integration to derive a boundary matching equation.
    The contour $\Gamma$ encloses and is curved along the boundary.
  }
  \label{fig:boundary_matching}
\end{figure}
We consider the surface integral over the area enclosed by a closed curve $\Gamma$ in the $xz$ plane as shown in \figref{fig:boundary_matching} so that the integral catches the geometry of the boundary and reflects it on the boundary matching equation.
In the limit that the enclosed area vanishes ($\delta h \rightarrow 0$),
from the Faraday's law,
we have
\begin{align}
  \lim_{\delta h \rightarrow 0}
  \oiint_\Gamma
  \mathrm{d}\vec{S} 
  \cdot 
  \nabla \times \vec{\mathcal{E}}_{\mathbf{x},z}
  &=
  -\lim_{\delta h \rightarrow 0}
  \oiint_\Gamma
  \mathrm{d}\vec{S} 
  \cdot 
  \frac{\partial}{\partial t} 
  \mu_0 \vec{\mathcal{H}}_{\mathbf{x},z},
  \\
  \lim_{\delta h \rightarrow 0}
  \oint_\Gamma
  \mathrm{d}\vec{r} 
  \cdot 
  \vec{\mathcal{E}}_{\mathbf{x},z} 
  &= 0,
\end{align}
where we have used the Stokes' theorem.
The vectorial line element $\mathrm{d} \vec{r}$ is given by means of the tangential vector $\vec{t}_1$ and a scalar element $\sqrt{1+{a_\mathbf{x}'}^2}\mathrm{d}x$,
and we can obtain an integral equation,
\begin{align}
  \int \mathrm{d}x \sqrt{1+{a_\mathbf{x}'}^2}  
  \vec{t}_1 
  \cdot 
  (\vec{\mathcal{E}}_{\mathbf{x},a_\mathbf{x}+0}
  - 
  \vec{\mathcal{E}}_{\mathbf{x},a_\mathbf{x}-0}) = 0,
  \label{eq:boundary_condition_pre}
\end{align}
where 
$\vec{\mathcal{E}}_{\mathbf{x},a_\mathbf{x}+0}
= \vec{\mathcal{E}}_{\mathbf{x},z}^\mathrm{inc} 
+ \vec{\mathcal{E}}_{\mathbf{x},z}^\mathrm{ref}$
and 
$\vec{\mathcal{E}}_{\mathbf{x},a_\mathbf{x}-0} 
= \vec{\mathcal{E}}_{\mathbf{x},z}^\mathrm{tra}$
are the electric fields evaluated above and below the boundary,
respectively.
Since Eq. \eqref{eq:boundary_condition_pre} is independent of the choice of the integration interval,
we can obtain a boundary matching equation for the electric field,
\begin{align}
  \eta 
  \vec{t}_1 
  \cdot
  (
  \vec{\mathcal{E}}_{\mathbf{x},a_\mathbf{x}}^\mathrm{inc} 
  +
  \vec{\mathcal{E}}_{\mathbf{x},a_\mathbf{x}}^\mathrm{ref}
  - 
  \vec{\mathcal{E}}_{\mathbf{x},a_\mathbf{x}}^\mathrm{tra}
  ) = 0.
  \label{eq:t_1_E}
\end{align}
This equation means that the tangential component of the field should be matched at each point of the boundary at each time.
Note that we set $\eta = \sqrt{1+{a_\mathbf{x}'}^2}$ for simplicity.

Let us consider the matching equation for the magnetic field.
Integrating the Amp\`{e}re-Maxwell equation,
\begin{align}
  \lim_{\delta h \rightarrow 0}
  \oiint_\Gamma
  \mathrm{d}\vec{S} 
  \cdot 
  \nabla \times \vec{\mathcal{H}}_{\mathbf{x},z}
  &=
  \lim_{\delta h \rightarrow 0}
  \oiint_\Gamma
  \mathrm{d}\vec{S} 
  \cdot 
  \frac{\partial}{\partial t} 
  \epsilon_{\mathbf{x},z}\epsilon_0
  \vec{\mathcal{E}}_{\mathbf{x},z},
  \\
  \lim_{\delta h \rightarrow 0}
  \oint_\Gamma
  \mathrm{d}\vec{r} 
  \cdot 
  \vec{\mathcal{H}}_{\mathbf{x},z}
  &=
  \frac{\dot{a}_{\mathbf{x}}}{c}
  \alpha
  \vec{t}_2
  \cdot
  \frac{\vec{\mathcal{E}}_{\mathbf{x},a_\mathbf{x}}^\mathrm{tra}}{Z_0},
  \label{eq:Ampere-Maxwell}
\end{align}
where we have used the Stokes' theorem and the time derivative of the permittivity distribution,
$
\partial \epsilon_{\mathbf{x},z}/\partial t 
=
\dot{a}_\mathbf{x}
\alpha
\delta(a_\mathbf{x}-z)
$.
Note that there remains a surface electric current on the right hand side of Eq. \eqref{eq:Ampere-Maxwell}.
This is because we induced electric polarisation on the boundary due to the permittivity difference $\alpha$.
We can write the matching equation for the magnetic field,
\begin{align}
  \eta
  \vec{t}_1 
  \cdot 
  (
  \vec{\mathcal{H}}_{\mathbf{x},a_\mathbf{x}}^\mathrm{inc} 
  +
  \vec{\mathcal{H}}_{\mathbf{x},a_\mathbf{x}}^\mathrm{ref}
  - 
  \vec{\mathcal{H}}_{\mathbf{x},a_\mathbf{x}}^\mathrm{tra}
  ) 
  =
  \frac{\dot{a}_{\mathbf{x}}}{c}
  \alpha
  \vec{t}_2
  \cdot
  \frac{\vec{\mathcal{E}}_{\mathbf{x},a_\mathbf{x}}^\mathrm{tra}}{Z_0}.
  \label{eq:t_1_H}
\end{align}

Following the same procedure,
we can write the boundary matching equations in the $y$ direction,
\begin{align}
  &\vec{t}_2 \cdot 
  (
  \vec{\mathcal{E}}{}_{\mathbf{x},a_\mathbf{x}}^\mathrm{inc} 
  + 
  \vec{\mathcal{E}}{}_{\mathbf{x},a_\mathbf{x}}^\mathrm{ref} 
  - 
  \vec{\mathcal{E}}{}_{\mathbf{x},a_\mathbf{x}}^\mathrm{tra}
  ) 
  = 0,
  \vspace{.5em}
  \label{eq:t_2_E}
  \\
  &\vec{t}_2 \cdot 
  (
  \vec{\mathcal{H}}{}_{\mathbf{x},a_\mathbf{x}}^\mathrm{inc} 
  +
  \vec{\mathcal{H}}{}_{\mathbf{x},a_\mathbf{x}}^\mathrm{ref} 
  - 
  \vec{\mathcal{H}}{}_{\mathbf{x},a_\mathbf{x}}^\mathrm{tra}
  )
  =
  -\frac{\dot{a}_\mathbf{x}}{c}
  \alpha
  \eta
  \vec{t}_1
  \cdot
  \frac{
    \vec{\mathcal{E}}{}_{\mathbf{x},a_\mathbf{x}}^\mathrm{tra}
  }{Z_0}.
  \label{eq:t_2_H}
\end{align}

Since our modulation is invariant with respect to the translation in the $y$ direction,
the problem can be regarded as a two-dimensional one.
Here, we investigate two fundamental cases where the incident field is either $s$- or $p$-polarised.

There is no polarisation rotation as we stick to the in-plane calculation.
This is because our `optical axis' induced by the surface structure of grating type is in the $y$ direction,
which is parallel or perpendicular to the electric and magnetic fields.
In the $s$ polarisation case,
the electric field oscillates perpendicularly to the $xz$ plane while the magnetic field oscillates in the $xz$ plane.
Therefore,
Eqs. (\ref{eq:t_1_E}, \ref{eq:t_2_H}) are automatically satisfied in the $s$ polarisation case,
and we can focus on
\begin{align}
  \begin{cases}{}
    \vec{t}_2 \cdot 
    (
    \vec{\mathcal{E}}{}_{\mathbf{x},a_\mathbf{x}}^\mathrm{inc} 
    + 
    \vec{\mathcal{E}}{}_{\mathbf{x},a_\mathbf{x}}^\mathrm{ref} 
    - 
    \vec{\mathcal{E}}{}_{\mathbf{x},a_\mathbf{x}}^\mathrm{tra}
    ) 
    = 0,
    \vspace{.5em}
    \\
    \eta
    \vec{t}_1 \cdot 
    Z_0 (
    \vec{\mathcal{H}}{}_{\mathbf{x},a_\mathbf{x}}^\mathrm{inc} 
    + 
    \vec{\mathcal{H}}{}_{\mathbf{x},a_\mathbf{x}}^\mathrm{ref} 
    -
    \vec{\mathcal{H}}{}_{\mathbf{x},a_\mathbf{x}}^\mathrm{tra}
    )
    =
    \displaystyle{
      \frac{\dot{a}_{\mathbf{x}}}{c}
    }
    \alpha
    \vec{t}_2
    \cdot
    \vec{\mathcal{E}}_{\mathbf{x},a_\mathbf{x}}^\mathrm{tra}.
  \end{cases}
  \label{eq:boundary_conditions_s_pol}
\end{align}

On the other hand,
the magnetic field oscillates in the $y$ direction in the $p$ polarisation case,
and the electric field lies in the $xz$ plane.
Thus, we can focus on
\begin{align}
  \begin{cases}{}
    \eta
    \vec{t}_1 \cdot 
    (
    \vec{\mathcal{E}}{}_{\mathbf{x},a_\mathbf{x}}^\mathrm{inc} 
    + 
    \vec{\mathcal{E}}{}_{\mathbf{x},a_\mathbf{x}}^\mathrm{ref} 
    -
    \vec{\mathcal{E}}{}_{\mathbf{x},a_\mathbf{x}}^\mathrm{tra}
    ) = 0,
    \vspace{.5em}
    \\
    \vec{t}_2 \cdot 
    Z_0 (
    \vec{\mathcal{H}}{}_{\mathbf{x},a_\mathbf{x}}^\mathrm{inc} 
    +
    \vec{\mathcal{H}}{}_{\mathbf{x},a_\mathbf{x}}^\mathrm{ref} 
    - 
    \vec{\mathcal{H}}{}_{\mathbf{x},a_\mathbf{x}}^\mathrm{tra}
    )
    =
    -\displaystyle{
      \frac{
        \dot{a}_\mathbf{x}  
      }{c}
    }
    \alpha
    \eta\vec{t}_1
    \cdot
    \vec{\mathcal{E}}{}_{\mathbf{x},a_\mathbf{x}}^\mathrm{tra}.
  \end{cases}
  \label{eq:boundary_conditions_p_pol}
\end{align}
These simultaneous equations depend on space and time.
Thanks to the time derivative of the surface profile $\dot{a}_\mathbf{x}$ and the tangential vector $\eta \vec{t}_1$ in Eqs. (\ref{eq:boundary_conditions_s_pol}, \ref{eq:boundary_conditions_p_pol}),
we can properly take both temporal and spatial modulations into consideration.

We apply the Fourier transform 
$
\mathscr{F}[g_\mathbf{x}]_\mathbf{k} 
\equiv 
\int  
g_\mathbf{x}
e^{-i\mathbf{k}\cdot\mathbf{x}}
d\mathbf{x}
$
to Eqs. (\ref{eq:boundary_conditions_s_pol}, \ref{eq:boundary_conditions_p_pol}) to obtain equation systems determining the reflection and transmission matrices in the frequency domain.

Let take the $s$ polarisation as an example.
We use the Fourier expansion of the electric field (\ref{eq:E,H_lnc,ref,tra}, \ref{eq:E,H_z,alpha^sigmatau}) to get
\begin{align}
  \vec{t}_2 
  \cdot 
  \vec{\mathcal{E}}_{\mathbf{x},a_\mathbf{x}}^{\Lambda}
  &=
  -\int_\mathbf{k}
  e^{i\mathbf{k}\cdot\mathbf{x}}
  \operatorname{sgn}(\omega)
  \frac{k_x}{k_\parallel}
  e^{i\phi_{\mathbf{k}}^{\sigma\tau}\sin \mathbf{q}\cdot\mathbf{x}}
  E_{s,\mathbf{k}}^{\sigma\tau},
\end{align}
where we have defined the propagating phase factor
$
\phi_{\mathbf{k}}^{\sigma \tau}
=\sigma K_{\mathbf{k}}^\tau A
$.
We use the Jacobi-Anger identity to expand the exponential of the trigonometric function 
\cite{cuyt2008handbook}, 
\begin{align}
  \vec{t}_2 
  \cdot 
  \vec{\mathcal{E}}_{\mathbf{x},a_\mathbf{x}}^{\Lambda}
  &=
  -\SumInt_{m,\mathbf{k}}
  e^{i\mathbf{k}\cdot\mathbf{x}}
  \frac{k_{x,m}}{k_{\parallel,m}}
  \operatorname{sgn}(\omega_m)
  J_{-m}(\phi_{\mathbf{k}_m}^{\sigma\tau})
  E_{s,\mathbf{k}_m}^{\sigma\tau},
\end{align}
where $J_m$ is the $m^\mathrm{th}$ order Bessel function of the first kind.
Note that the integration variables are relevantly shifted,
$\mathbf{k} \mapsto \mathbf{k}_m = \mathbf{k}+m\mathbf{q}$,
after the expansion.
Note also that we have introduced a shorthand notation,
\begin{align}
  \SumInt_{m,\mathbf{k}}
  = 
  \sum_{m=-\infty}^{+\infty}
  \int_\mathbf{k}.
\end{align}

Applying the Fourier transform,
we can obtain the $l$th order quantity,
\begin{align}
  \mathscr{F}
  \left[
    \vec{t}_2 
    \cdot 
    \vec{\mathcal{E}}{}_{\mathbf{x},a_\mathbf{x}}^{\Lambda} 
  \right]_{\mathbf{k}_l}
  =
  \left[
    \mathsf{M}_{\mathbf{k}}^{\sigma\tau}
  \mathbb{E}_{s,\mathbf{k}}^{\sigma\tau}
\right]_l.
\end{align}
Here, we have collected the modal amplitude in each diffraction order into one column,
\begin{align}
  \mathbb{E}_{\lambda,\mathbf{k}}^{\sigma\tau}
   &=
   \begin{pmatrix}
     \vdots
     \\
     E_{\lambda,\mathbf{k}_{-1}}^{\sigma\tau}
     \\
     E_{\lambda,\mathbf{k}_{0}}^{\sigma\tau}
     \\
     E_{\lambda,\mathbf{k}_{+1}}^{\sigma\tau}
     \\
     \vdots
   \end{pmatrix},
   \label{eq:modal_amp_vec}
\end{align}
and introduced a matrix-vector representation with the coefficient matrix that has the geometric information of the boundary,
\begin{align}
  [\mathsf{M}_{\mathbf{k}}^{\sigma\tau}]_{lm}
  &=
  \frac{k_{x,m}}{k_{\parallel,m}}
  \operatorname{sgn}(\omega_m)
  J_{l-m}(\phi_{\mathbf{k}_m}^{\sigma\tau}).
  \label{eq:M}
\end{align}
Here, the Bessel function is responsible for the correlation between the $l$th order and the $m$th order diffraction.
The diffraction is stronger as the surface corrugation depth increases with respect to the wavenumber in the $z$ direction.
This is why we provide $\phi_{\mathbf{k}_m}^{\sigma \tau} = \sigma K_{\mathbf{k}_m}^{\sigma \tau} A$ in the argument.

Finally,
applying the Fourier transform to Eq. \eqref{eq:t_2_E},
we can obtain
\begin{align}
  \mathsf{M}_{\mathbf{k}}^{\Inc}
  \mathbb{E}_{s,\mathbf{k}}^{\Inc}
  +
  \mathsf{M}_{\mathbf{k}}^{\Refl}
  \mathbb{E}_{s,\mathbf{k}}^{\Refl}
  -
  \mathsf{M}_{\mathbf{k}}^{\Tra}
  \mathbb{E}_{s,\mathbf{k}}^{\Tra}
  &= 0.
  \label{eq:t_2_E_FT}
\end{align}

Similarly,
we can perform the Fourier transform for the boundary condition of the magnetic field.
The tangential component of the magnetic field is evaluated as below:
\begin{align}
  &\eta
  \vec{t}_1
  \cdot
  Z_0
  \vec{\mathcal{H}}_{\mathbf{x},a_\mathbf{x}}^{\Lambda}
  =
  \int_\mathbf{k}
  e^{i\mathbf{k}\cdot\mathbf{x}}
  e^{
    i\sigma K_{\mathbf{k}}^\tau a_\mathbf{x}
  }
  \eta
  \vec{t}_1
  \cdot
  \vec{h}_{s,\mathbf{k}}^{\sigma\tau}
  Z_0
  H_{s,\mathbf{k}}^{\sigma\tau},
  \\
  &=
  \int_\mathbf{k}
  e^{i\mathbf{k}\cdot\mathbf{x}} 
  \frac{\sigma K_\mathbf{k}^\tau k_x - a_\mathbf{x}' {k_\parallel}^2}
  {k_\parallel |k_0|}
  e^{
    i\sigma 
    K_{\mathbf{k}}^\tau
    a_\mathbf{x}
  }
  E_{s,\mathbf{k}}^{\sigma \tau},
  \\
  &=
  \SumInt_{m,\mathbf{k}}
  e^{
    i(\mathbf{k} - m\mathbf{q})
    \cdot\mathbf{x}
  }
  \left(
    \frac{k_x}
    {k_\parallel}
    \frac{\sigma K_\mathbf{k}^\tau}
    {|k_0|}
    -
    \frac{-mg}
    {\sigma K_{\mathbf{k}}^\tau}
    \frac{k_\parallel}
    {|k_0|}
  \right)
  J_{-m}(\phi_{\mathbf{k}}^{\sigma\tau}) 
  E_{s,\mathbf{k}}^{\sigma \tau},
\end{align}
where we expanded the exponential of the trigonometric function by using the Bessel functions.
Applying the Fourier transform to Eq. \eqref{eq:t_1_H} gives
\begin{align}
  \mathsf{N}_{\mathbf{k}}^{\Inc}
  \mathbb{E}_{s,\mathbf{k}}^{\Inc}
  +
  \mathsf{N}_{\mathbf{k}}^{\Refl}
  \mathbb{E}_{s,\mathbf{k}}^{\Refl}
  -
  (
  \mathsf{N}_{\mathbf{k}}^{\Tra}
  +
  \mathsf{L}_{\mathbf{k}}
  )
  \mathbb{E}_{s,\mathbf{k}}^{\Tra}
  = 0.
  \label{eq:t_1_H_FT}
\end{align}
Here,
the element of the coefficient matrix is given by 
\begin{align}
  [\mathsf{L}_{\mathbf{k}}]_{lm}
  &=
  \frac{\alpha}{c}
  \frac{(l-m)\Omega}{-K_{\mathbf{k}_m}^{\ssl}}
  \times
  \operatorname{sgn}(\omega_m)
  \frac{k_{x,m}}{k_{\parallel,m}}
  J_{l-m}(\phi_{\mathbf{k}_m}^{\ssm\ssl}),
  \label{eq:L}
  \\
  [\mathsf{N}_{\mathbf{k}}^{\sigma\tau}]_{lm}
  &=
  \left(
    \frac{k_{x,m}}{k_{\parallel,m}}
    \frac{\sigma K_{\mathbf{k}_m}^\tau}{|k_{0,m}|}
    -
    \frac{(l-m)g}{\sigma K_{\mathbf{k}_m}^\tau}
    \frac{k_{\parallel,m}}{|k_{0,m}|}
  \right)
  J_{l-m}(\phi_{\mathbf{k}_m}^{\sigma\tau}).
  \label{eq:N}
\end{align}
The expressions of $\mathsf{L}$ and $\mathsf{N}$ are relatively intricate,
compared with the $\mathsf{M}$ matrix.
The $\mathsf{L}$ matrix is responsible for the induced surface current \eqref{eq:Ampere-Maxwell},
and thus its matrix elements are proportional to the permittivity difference $\alpha$.
The factor of $[(l-m)\Omega]/(-K_{\mathbf{k}_m}^{\ssl})$ is produced by the time derivative $\dot{a}_\mathbf{x}$ that we have in the expression of the surface current \eqref{eq:Ampere-Maxwell}.
As for the $\mathsf{N}$ matrix,
the non-uniform vectorial line element,
$
\eta
\vec{t}_1 
=
\vec{u}_x + a_\mathbf{x}'\vec{u}_z
$,
makes the expression complicated
The first and second terms in Eq. \eqref{eq:N} stems from the field matching in the $x$ and $z$ directions, respectively.
The space derivative $a_\mathbf{x}'$ yields the factor of $[(l-m)g]/(\sigma K_{\mathbf{k}_m}^\tau)$ in the second term.
It is worth noting that here we can define the incident angle $\theta_\mathrm{in}$ by
\begin{align}
  \cos \theta_\mathrm{in}
  &= \frac{\sigma K_{\mathbf{k}}^{\ssg}}{|k_{0}|},
  \quad
  \sin \theta_\mathrm{in}
  = \frac{k_{\parallel}}{|k_{0}|}.
  \label{eq:theta_in}
\end{align}

Here, we observe that the $\mathsf{M}$, $\mathsf{N}$ and $\mathsf{L}$ matrices does not change if we keep the ratio between any two lengths and just scale the parameters.
All of the matrix elements are given by means of dimensionless numbers such as $ K_{\mathbf{k}}^< A$ and $g/K_{\mathbf{k}}^<$.

Those quantities are invariant under the scaling of the modulation depth $A$ and the reciprocal vectors,
$\mathbf{q}$ and $\mathbf{k}$,
\begin{align}
  \begin{cases}{}
    A \mapsto \beta A,
    \quad
    \mathbf{q} \mapsto \beta^{-1} \mathbf{q},
    \\
    \mathbf{k} \mapsto \beta^{-1} \mathbf{k},
  \end{cases}
\end{align}
where $\beta$ is the scale factor.
Since the effects of the spatiotemporal modulation is encoded by the $\mathsf{M}$, $\mathsf{N}$ and $\mathsf{L}$ matrices,
the scaling does not affect the calculation.

Let us rearrange Eqs. (\ref{eq:t_2_E_FT}, \ref{eq:t_1_H_FT}) in a matrix form,
\begin{align}
  \begin{pmatrix}
    \mathsf{M}_{\mathbf{k}}^{\Refl} 
    &
    -\mathsf{M}_{\mathbf{k}}^{\Tra}
    \\
    \mathsf{N}_{\mathbf{k}}^{\Refl} 
    &
    -(\mathsf{N}_{\mathbf{k}}^{\Tra} 
    +
    \mathsf{L}_{\mathbf{k}} 
    )
  \end{pmatrix}
  \begin{pmatrix}
    \mathbb{E}_{s,\mathbf{k}}^{\Refl}
    \\
    \mathbb{E}_{s,\mathbf{k}}^{\Tra}
  \end{pmatrix}
  = 
  \begin{pmatrix}
    -\mathsf{M}_{\mathbf{k}}^{\Inc}
    \mathbb{E}_{s,\mathbf{k}}^{\Inc}
    \\
    -\mathsf{N}_{\mathbf{k}}^{\Inc} 
    \mathbb{E}_{s,\mathbf{k}}^{\Inc}
  \end{pmatrix}.
  \label{eq:s_pol}
\end{align}
This is the scattering equation for the $s$ polarisation incidence.

By following the same procedure,
we can obtain the matrix equation for the $p$ polarisation,
\begin{align}
  \begin{pmatrix}
    \mathsf{M}_{\mathbf{k}}^{\Refl} 
    &
    -(\mathsf{M}_{\mathbf{k}}^{\Tra}
    +\widetilde{\mathsf{L}}_{\mathbf{k}})
    \\
    \widetilde{\mathsf{N}}_{\mathbf{k}}^{\Refl} 
    &
    -\widetilde{\mathsf{N}}_{\mathbf{k}}^{\Tra} 
  \end{pmatrix}
  \begin{pmatrix}
    \mathbb{H}_{p,\mathbf{k}}^{\Refl}
    \\
    \mathbb{H}_{p,\mathbf{k}}^{\Tra}
  \end{pmatrix}
  = 
  \begin{pmatrix}
    -\mathsf{M}_{\mathbf{k}}^{\Inc} 
    \mathbb{H}_{p,\mathbf{k}}^{\Inc}
    \\
    -\widetilde{\mathsf{N}}_{\mathbf{k}}^{\Inc} 
    \mathbb{H}_{p,\mathbf{k}}^{\Inc}
  \end{pmatrix},
  \label{eq:p_pol}
\end{align}
where we have defined
$
\widetilde{\mathsf{N}}_{\mathbf{k}}^{\sigma\tau} 
=
\mathsf{N}_{\mathbf{k}}^{\sigma\tau}/\epsilon^\tau
$.
Note that the electric current contribution $\widetilde{\mathsf{L}}$ appears not with the $\widetilde{\mathsf{N}}$ matrix but with the $\widetilde{\mathsf{M}}$ matrix.
Please see the Appendix \ref{app:p_pol} for the derivation.

\section{Numerical implementation}
\label{sec:numerial}
Inverting Eqs. (\ref{eq:s_pol}, \ref{eq:p_pol}),
we can get the reflection and transmission matrices,
\begin{align}
  \begin{pmatrix}
    \mathsf{R}_{s,\mathbf{k}}
    \\
    \mathsf{T}_{s,\mathbf{k}}
  \end{pmatrix}
  &=
  \begin{pmatrix}
    \mathsf{M}_{\mathbf{k}}^{\Refl} 
    &
    -\mathsf{M}_{\mathbf{k}}^{\Tra}
    \\
    \mathsf{N}_{\mathbf{k}}^{\Refl} 
    &
    -(
    \mathsf{N}_{\mathbf{k}}^{\Tra} 
    +
    \mathsf{L}_{\mathbf{k}}
    )
  \end{pmatrix}^{-1}
  \begin{pmatrix}
    -\mathsf{M}_{\mathbf{k}}^{\Inc}
    \\
    -\mathsf{N}_{\mathbf{k}}^{\Inc}
  \end{pmatrix},
  \label{eq:RTs}
  \\
  \begin{pmatrix}
    \mathsf{R}_{p,\mathbf{k}}
    \\
    \mathsf{T}_{p,\mathbf{k}}
  \end{pmatrix}
  &=
  \begin{pmatrix}
    \mathsf{M}_{\mathbf{k}}^{\Refl} 
    &
    -(\mathsf{M}_{\mathbf{k}}^{\Tra}
    +\widetilde{\mathsf{L}}_{\mathbf{k}})
    \\
    \widetilde{\mathsf{N}}_{\mathbf{k}}^{\Refl} 
    &
    -\widetilde{\mathsf{N}}_{\mathbf{k}}^{\Tra} 
  \end{pmatrix}^{-1}
  \begin{pmatrix}
    -\mathsf{M}_{\mathbf{k}}^{\Inc}
    \\
    -\mathsf{N}_{\mathbf{k}}^{\Inc}
  \end{pmatrix}.
  \label{eq:RTp}
\end{align}
Note that only the diagonal elements of $\mathsf{M}_{\mathbf{k}}^{\sigma \tau}$ and $\mathsf{N}_{\mathbf{k}}^{\sigma \tau}$ remain finite in the flat boundary limit ($A\rightarrow 0$),
and all other matrix elements vanish so that Eqs. (\ref{eq:RTs}, \ref{eq:RTp}) recover the Fresnel coefficients
(See the Appendix \ref{app:Fresnel}).

\begin{figure}[htbp]
  \includegraphics[width=\linewidth]{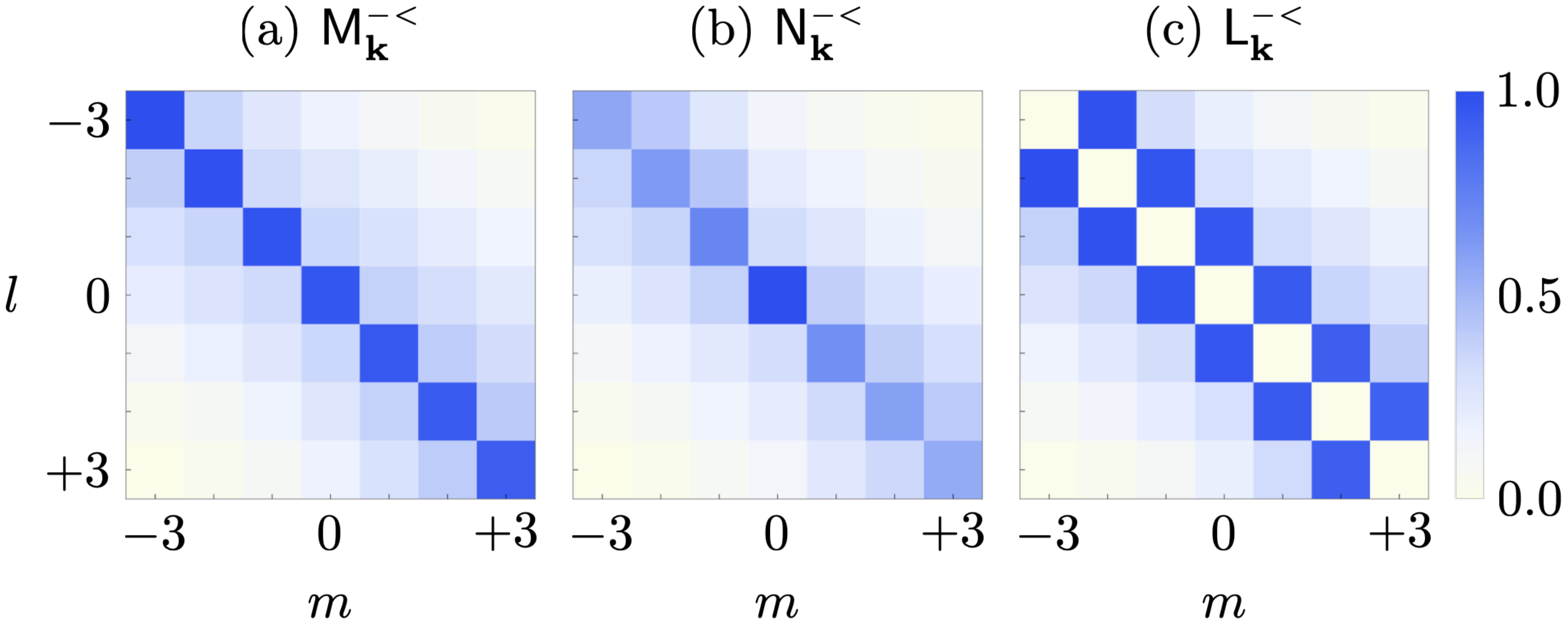}
  \caption{
    Typical distribution of the matrix elements.
    The absolute values of the matrix elements in $\mathsf{M}$, $\mathsf{N}$ and $\mathsf{L}$ are plotted.
    The common colorbar is shown on the right of (c).
    The horizontal and vertical axes are the column and row indices.
    Note that all elements are normalised by the maximum element in each matrix.
    In these plots,
    we use the following parameters:
    $\epsilon^{\ssg} = 1.0$,
    $\epsilon^{\ssl} = 2.25$,
    $A = 10\ \mathrm{[nm]}$,
    $g = 2\pi\ \mathrm{[\mu m^{-1}]}$,
    $\Omega = 0.2gc$,
    $\omega_\mathrm{in} = 0.8 gc$,
    $\theta_\mathrm{in} = 0$,
    $k_y = 0$.
    The cutoff number is $m_c = 3$.
  }
  \label{fig:decay}
\end{figure}

When we numerically evaluate the reflection and transmission matrices,
we truncate the $\mathsf{M}$, $\mathsf{N}$ and $\mathsf{L}$ matrices to finite rank ones 
so that 
$-m_c \leq l \leq +m_c$
and
$-m_c \leq m \leq +m_c$,
where $m_c$ is the cutoff number.
Since our boundary is differentiable and the media above and below the boundary are homogeneous,
the truncation can be justified as in the conventional Fourier modal methods
\cite{%
  li1999justification,
  shcherbakov2013efficient%
}.
As shown in \figref{fig:decay},
we can also numerically confirm that large matrix elements are highly localised near the diagonal elements.

\begin{figure}[htbp]
  \centering
  \includegraphics[width=\linewidth]{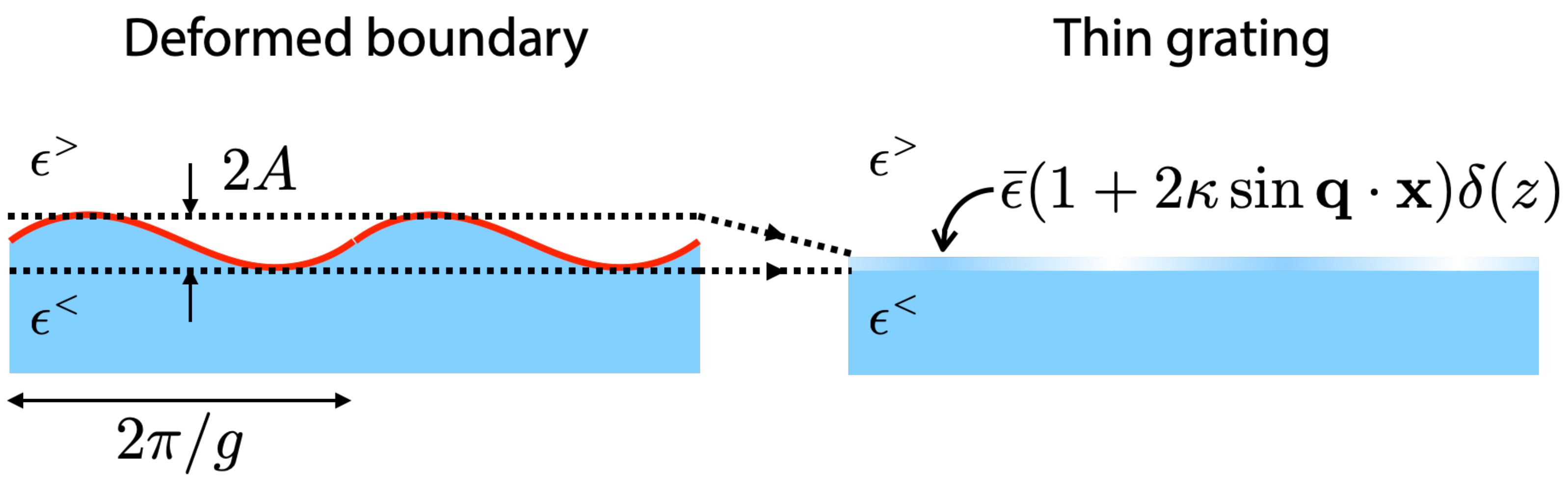}
  \caption{
    Homogenisation.
    Averaging the permittivity in the $z$ direction within the modulated region ($-A \leq z \leq +A$) generates a very thin grating model.
  }
  \label{fig:homogenisation}
\end{figure}

When the modulation strength is sufficiently weak compared with the spatial and temporal modulation period ($gA \ll 1$, and $\Omega A/c \ll 1$),
the surface structure is homogenised from the perspective of the electromagnetic field.
In other words,
the field is slowly varying near the grating,
and thus we can consider effective infinitesimally thin medium to model the deformed boundary.
The effective medium is given by means of permittivity averaging in the $z$ direction,
\begin{align}
  \epsilon_{\mathbf{x}}^\mathrm{sf}
  &= \int_{-A}^{+A} \mathrm{d}z\  
  \epsilon_{\mathbf{x},z}
  =
  \bar{\epsilon} (1 + 2i\kappa \sin \mathbf{q} \cdot \mathbf{x})
  \label{eq:bar_epsilon}
\end{align}
where we have defined effective parameters,
\begin{align}
  \bar{\epsilon} 
  &= (\epsilon^\ssl + \epsilon^\ssg)A,
  \quad
  \kappa 
  = \frac{1}{2i}\frac{\epsilon^\ssl - \epsilon^\ssg}{\epsilon^\ssl + \epsilon^\ssg}.
\end{align}
With the permittivity averaging procedure,
we can get a very thin grating as the effective thin medium (see \figref{fig:homogenisation}).
The scattering calculation within the effective medium description is shown in the Supplemental Materials
\footnote{
  See the Supplemental Materials.
}.

In \figref{fig:angular},
we compare the incident angle dependences of the first order diffraction amplitudes calculated within the dynamical differential formalism and that within the homogenisation theory.
We can clearly recognise that the angular spectra calculated in two different method agree well.
Under the flipping both of the sign of diffraction orders and the incident angle,
the graph is symmetric if there is no time dependence ($\Omega=0$).
This is a consequence of the discrete translational invariance of the system.
If we introduce the temporal modulation ($\Omega\neq 0$),
the system is no longer invariant the translation and nonreciprocal.
This is why the plots are asymmetric when $\Omega\neq0$.
\begin{figure}[]
  \centering
  \includegraphics[width=\linewidth]{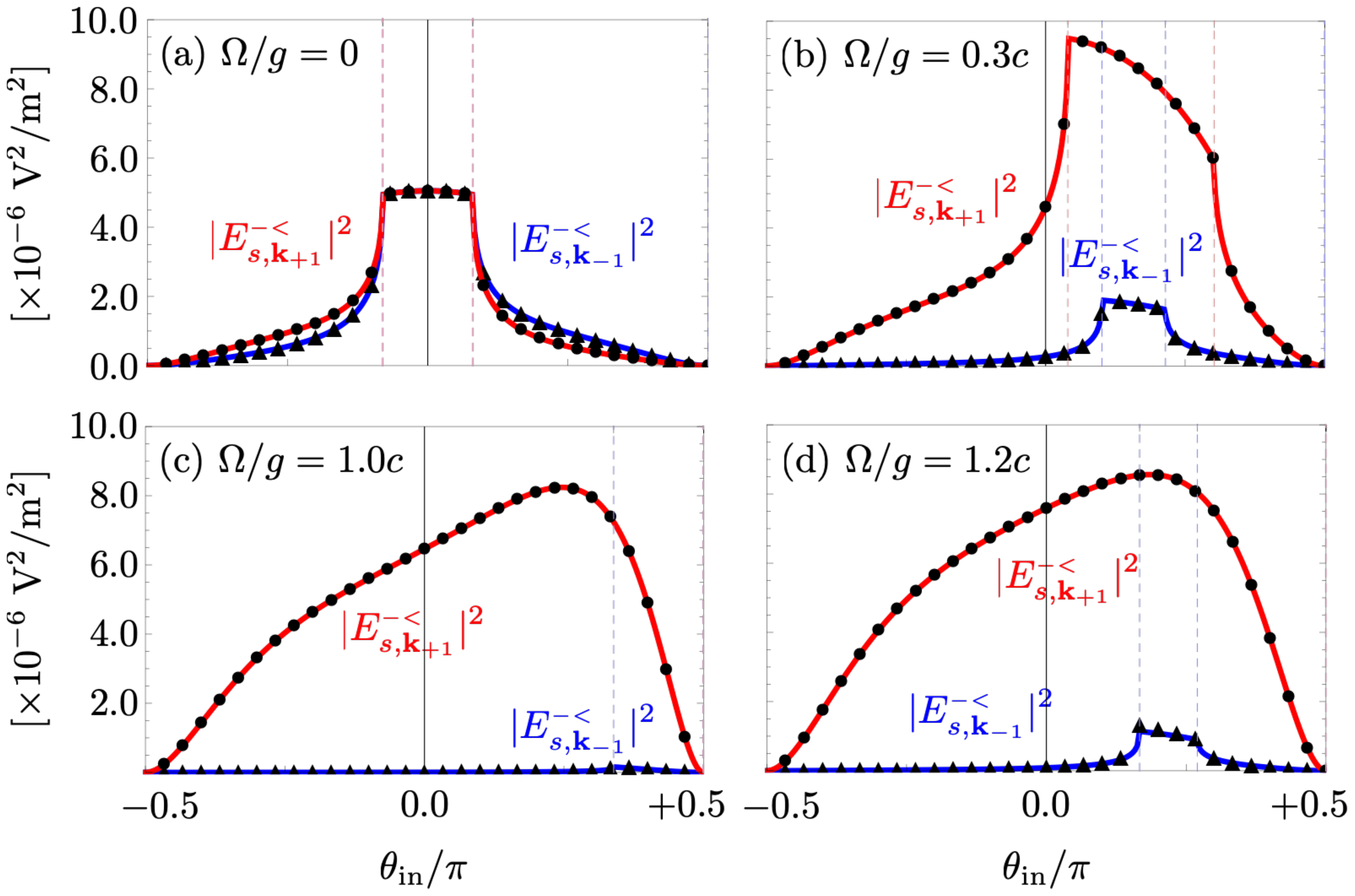}%
  \caption{
    Incident angle dependences of the first order diffraction intensities.
    The horizontal axes are the incident angle defined by Eq. \eqref{eq:theta_in}.
    The blue (red) curve corresponds to the negative (positive) diffraction $|E_{s,\mathbf{k}_{-1}}^{\ssm\ssl}|^2$ ($|E_{s,\mathbf{k}_{+1}}^{\ssm\ssl}|^2$) inside the dielectric medium calculated by the dynamical differential formalism.
    The black circles and triangles are generated by the effective medium description.
    At dashed lines corresponding to the emergence of diffraction modes
    ($K_{\mathbf{k}_{\pm 1}}^\tau=0$),
    the angular spectra are singular,
    that is a Wood grating anomaly \cite{wood1902xlii}.
    In these figures,
    we substitute the following parameters:
    $\epsilon^{\ssg} = 1.0$,
    $\epsilon^{\ssl} = 2.25$,
    $g = 2\pi\ \mathrm{[\mu m^{-1}]}$,
    $A = 1\ \mathrm{[nm]}$,
    $\lambda = s$,
    $\omega_\mathrm{in} = 0.8 gc$,
    $k_y = 0$.
    The cutoff number is $m_c = 3$.
  }
  \label{fig:angular}
\end{figure}

\begin{figure}[htbp]
  \centering
  \includegraphics[width=\linewidth]{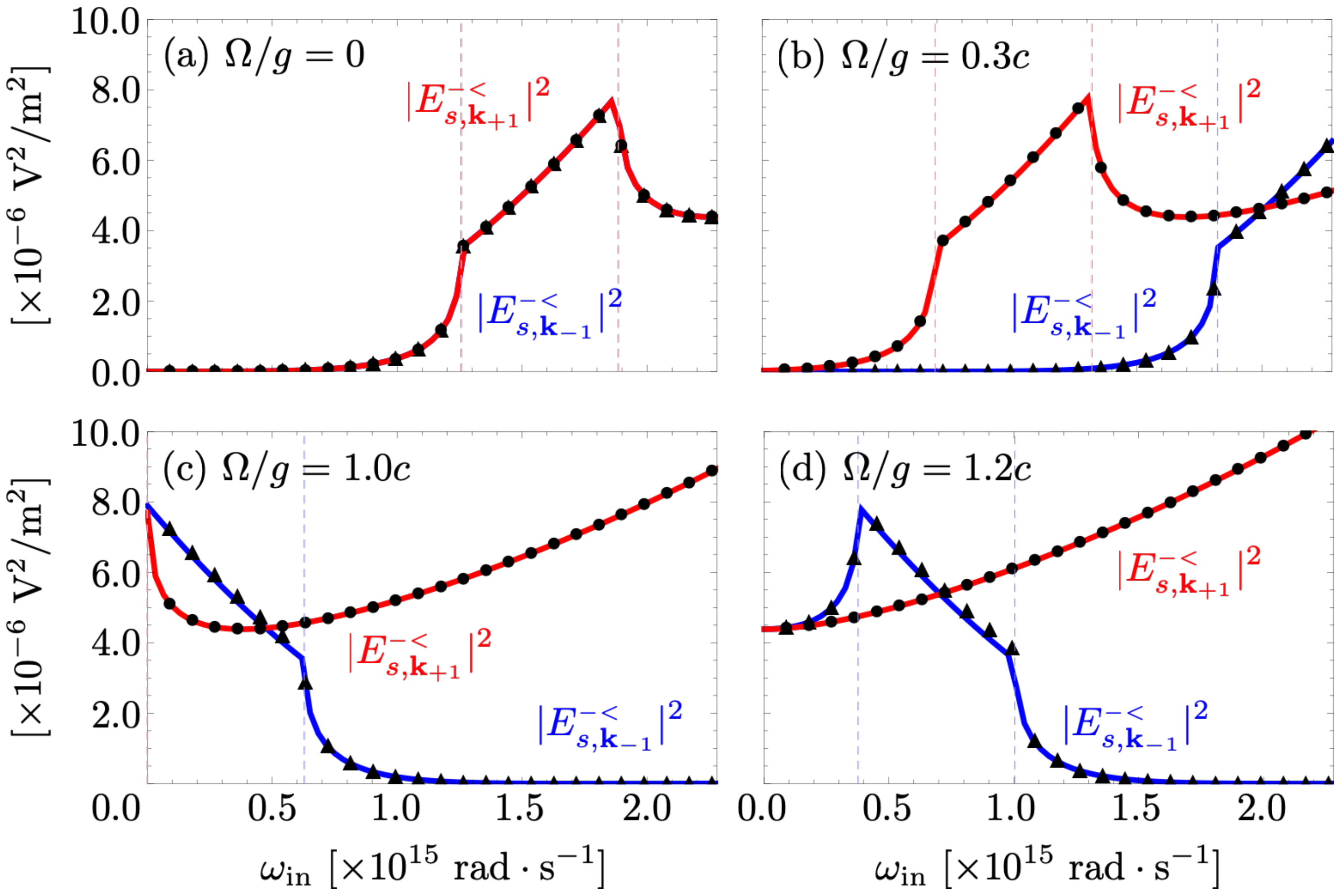}
  \caption{
    Frequency spectra of the first order diffraction amplitudes.
    The blue (red) curve corresponds to the negative (positive) diffraction amplitude $|E_{s,\mathbf{k}_{-1}}^{\ssm\ssl}|^2$ ($|E_{s,\mathbf{k}_{+1}}^{\ssm\ssl}|^2$) calculated by the dynamical differential formalism.
    Note that red and blue curves are completely overlap one another in (a).
    In these plots,
    we use the following parameters:
    $\epsilon^{\ssg} = 1.0$,
    $\epsilon^{\ssl} = 2.25$,
    $g = 2\pi\ \mathrm{[\mu m^{-1}]}$,
    $A = 1\ \mathrm{[nm]}$,
    $\lambda=s$,
    $\theta_\mathrm{in} = 0$,
    $k_y = 0$.
    The cutoff number is $m_c = 3$.
  }
  \label{fig:frequency}
\end{figure}

In \figref{fig:frequency},
we compare the first order diffraction amplitudes spectra generated by the dynamical differential formalism and the homogenisation approach.
We can see that the two approaches agree well.
If there is no temporal modulation ($\Omega = 0$),
the positive and negative diffraction spectra are the same.
This recovers the fact that diffraction at the static grating is symmetric.
Once the temporal modulation is introduced ($\Omega \neq 0$),
the positive and negative spectra start to deviate from each other.
Again, this implies that the spatiotemporal modulation breaks the reciprocity of our system.

\begin{figure}[htbp]
  \centering
  \includegraphics[width=\linewidth]{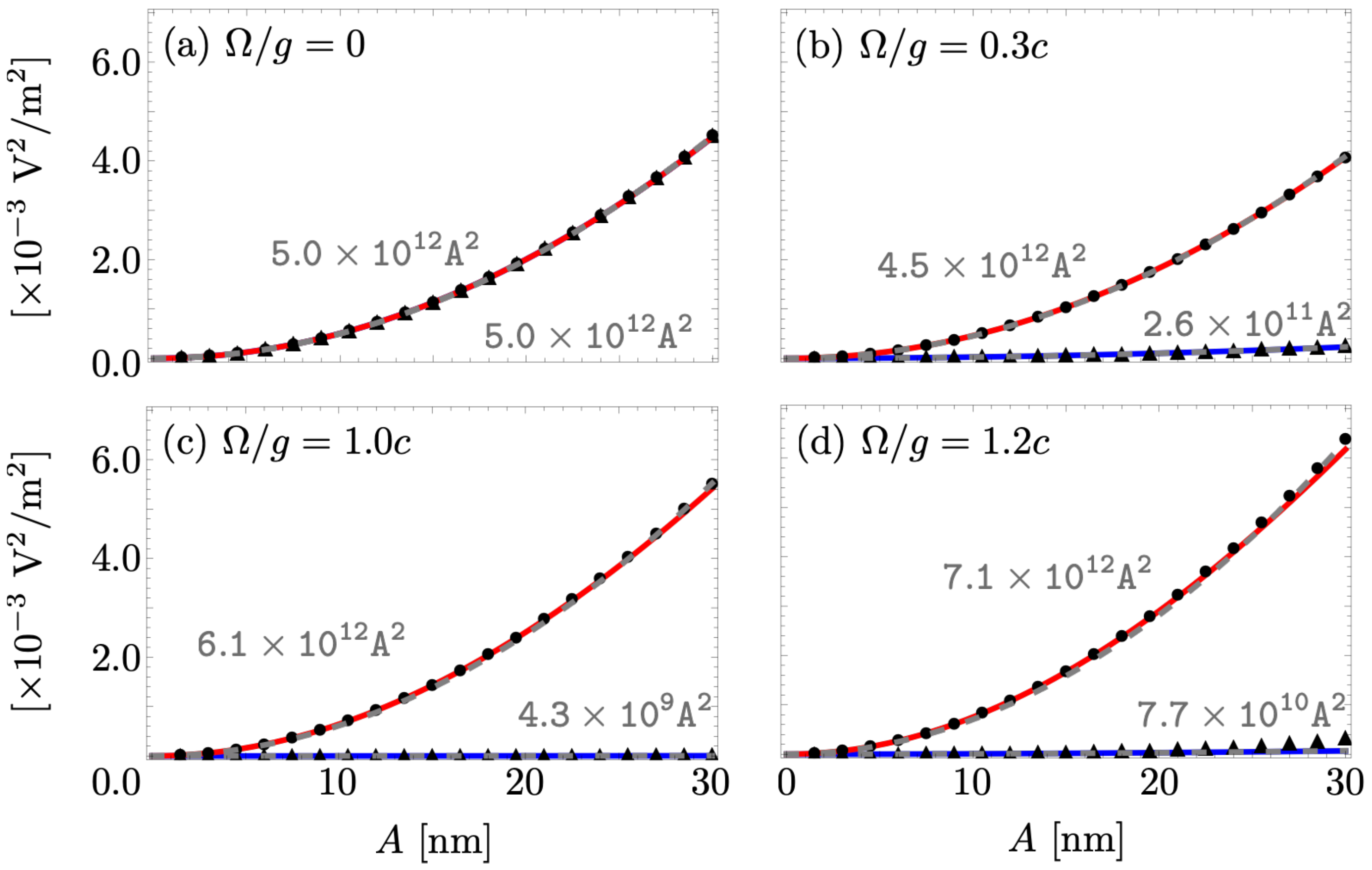}%
  \caption{
    Modulation strength dependence of the first order diffraction intensities.
    The blue (red) curve corresponds to the negative (positive) diffraction $|E_{s,\mathbf{k}_{-1}}^{\ssm\ssl}|^2$ ($|E_{s,\mathbf{k}_{+1}}^{\ssm\ssl}|^2$) inside the dielectric medium calculated by the dynamical differential formalism.
    The black circles and triangles are produced by the effective medium approach.
    Both positive and negative intensities quadratically depend on the modulation strength as the data is fitted by parabollic curves (grey dashed curves).
    The fitting equations are shown in each figure.
    Note that positive and negative diffraction intensities are completely overlap one another in (a).
    In these plots,
    we use the following parameters:
    $\epsilon^{\ssg} = 1.0$,
    $\epsilon^{\ssl} = 2.25$,
    $g = 2\pi\ \mathrm{[\mu m^{-1}]}$,
    $\omega_\mathrm{in} = 0.8 gc$,
    $\theta_\mathrm{in} = 0$,
    $k_y = 0$.
    The cutoff number is $m_c = 3$.
  }
  \label{fig:amplitude}
\end{figure}

In \figref{fig:amplitude},
we compare the modulation strength dependences of the first order diffraction amplitudes.
It is clear that two approaches agree when the modulation strength is small.
Both positive and negative diffraction intensities have quadratic dependence.

\section{Conclusions}
\label{sec:conclusion}
In this study,
we proposed a dynamical differential formalism which enables analytical calculation of the scattering by a surface modulated both in space and time.
Using dynamical coordinate transformation generated by the boundary profile,
we can directly impose the boundary conditions at the dynamically deformed interface for both electric and magnetic fields to properly capture the geometry and motion.
In the numerical calculation,
we confirmed our formalism is consistent with the effective medium description.
The diffraction spectra of the surface become asymmetric in the presence temporal modulation.

\begin{acknowledgments}
  D.O. is funded by the President's PhD Scholarships at Imperial College London.
  K.D. and J.B.P. acknowledges support from the Gordon and Betty Moore Foundation.
\end{acknowledgments}

\appendix
\section{\texorpdfstring{$p$}{p} polarisation case}
\label{app:p_pol}
In the $p$ polarisation case,
the tangential component of the electric field is
\begin{align}
  \eta
  \vec{t}_1
  \cdot
  \vec{\mathcal{E}}_{\mathbf{x},a_\mathbf{x}}^{\Lambda}
  &=
  \int_\mathbf{k}
  e^{i\mathbf{k}\cdot\mathbf{x}}
  e^{i\sigma K_\mathbf{k}^\tau a_\mathbf{x}}
  Z_{p,\mathbf{k}}^\tau
  H_{p,\mathbf{k}}^{\sigma \tau}
  \eta \vec{t}_1
  \cdot
  \vec{e}_{p,\mathbf{k}}^{\hspace{.2em}\sigma\tau}
  \\
  &=
  \int_\mathbf{k}
  e^{i\mathbf{k}\cdot\mathbf{x}}
  \frac{Z_0}{\epsilon^\tau}
  e^{i\sigma K_\mathbf{k}^\tau a_\mathbf{x}}
  \frac{\sigma K_\mathbf{k}^\tau k_x - a_\mathbf{x}' {k_\parallel}^2}
  {k_\parallel |k_0|}
  H_{p,\mathbf{k}}^{\sigma \tau},
\end{align}
where we have used the fact that the electric and magnetic amplitude can be associated with each other,
\begin{align}
  \begin{cases}{}
    E_{s,\mathbf{k}}^{\sigma \tau}
    =
    Z_{s,\mathbf{k}}^\tau
    H_{s,\mathbf{k}}^{\sigma \tau},
    \vspace{.5em}
    \\
    E_{p,\mathbf{k}}^{\sigma \tau}
    =
    Z_{p,\mathbf{k}}^\tau
    H_{p,\mathbf{k}}^{\sigma \tau},
  \end{cases}
\end{align}
via the characteristic impedance,
\begin{align}
  Z_{\lambda,\mathbf{k}}^\tau
  =
  \begin{cases}{}
    \cfrac{Z_0}{\sqrt{\epsilon^\tau}}
    \sqrt{\cfrac{\omega^2\epsilon^\tau/c^2}{|K_{\mathbf{k}}^\tau|^2 + {k_\parallel}^2}}
    &
    (\lambda = s),
    \vspace{0.5em}
    \\
    \cfrac{Z_0}{\sqrt{\epsilon^\tau}}
    \sqrt{\cfrac{|K_{\mathbf{k}}^\tau|^2 + {k_\parallel}^2}{\omega^2\epsilon^\tau/c^2}}
    &
    (\lambda = p),
  \end{cases}
\end{align}
where $Z_0=\sqrt{\mu_0/\epsilon_0}$ is the impedance of free space.
We use the Jacobi-Anger identity and perform the Fourier transform,
\begin{align}
  \mathscr{F}
  \left[
    \eta
    \vec{t}_1
    \cdot
    \vec{\mathcal{E}}_{\mathbf{x},z}^{\Lambda}
  \right]_{\mathbf{k}_l}
  &=
  Z_0
  \left[
    \widetilde{\mathsf{N}}_{\mathbf{k}}^{\sigma\tau}
    \mathbb{H}_{p,\mathbf{k}}^{\sigma\tau}
  \right]_l,
\end{align}
where we have defined 
$\widetilde{\mathsf{N}}_{\mathbf{k}}^{\sigma\tau} = \mathsf{N}_{\mathbf{k}}^{\sigma\tau}/\epsilon^\tau$.
Applying the Fourier transform to Eq. \eqref{eq:t_1_E} gives
\begin{align}
  \widetilde{\mathsf{N}}_{\mathbf{k}}^{\Inc}
  \mathbb{H}_{p,\mathbf{k}}^{\Inc}
  +
  \widetilde{\mathsf{N}}_{\mathbf{k}}^{\Refl}
  \mathbb{H}_{p,\mathbf{k}}^{\Refl}
  -
  \widetilde{\mathsf{N}}_{\mathbf{k}}^{\Tra}
  \mathbb{H}_{p,\mathbf{k}}^{\Tra}
  &= 0.
\end{align}
Note that we have divide the equation by the impedance of free space $Z_0$ to make the coefficient matrix dimensionless.

Next, we evaluate the tangential component of the magnetic field,
\begin{align}
  \vec{t}_2 
  \cdot 
  \vec{\mathcal{H}}_{\mathbf{x},a_\mathbf{x}}^{\Lambda}
  &=
  -\int_\mathbf{k}
  e^{i\mathbf{k}\cdot\mathbf{x}}
  \operatorname{sgn}(\omega)
  \frac{k_x}{k_\parallel}
  e^{i\phi_{\mathbf{k}}^{\sigma\tau}\sin \mathbf{q}\cdot\mathbf{x}}
  H_{p,\mathbf{k}}^{\sigma\tau}.
\end{align}
Using the Jacobi-Anger identity and applying the Fourier transform,
we can obtain the $l$th order quantity,
\begin{align}
  \mathscr{F}
  \left[
    \vec{t}_2 
    \cdot 
    \vec{\mathcal{H}}{}_{\mathbf{x},a_\mathbf{x}}^{\Lambda} 
  \right]_{\mathbf{k}_l}
  &=
  \left[
    \mathsf{M}_{\mathbf{k}}^{\sigma\tau}
    \mathbb{H}_{p,\mathbf{k}}^{\sigma\tau}
  \right]_l,
\end{align}
where the $\mathsf{M}$ matrix is given in Eq. \eqref{eq:M}.

The surface current contribution is evaluated as
\begin{align}
  \mathscr{F}
  \left[
    -\displaystyle{
      \frac{
        \dot{a}_\mathbf{x}  
      }{c}
    }
    \alpha
    \eta\vec{t}_1
    \cdot
    \frac{
      \vec{\mathcal{E}}{}_{\mathbf{x},a_\mathbf{x}}^\mathrm{tra}
    }{Z_0}
  \right]_{\mathbf{k}_l}
  &=
  \left[
    \widetilde{\mathsf{L}}_{\mathbf{k}}
    \mathbb{H}_{p,\mathbf{k}}^{\Tra}
  \right]_l
\end{align}
where we have introduced 
\begin{align}
  \widetilde{\mathsf{L}}_\mathbf{k}
  &=
  \frac{A\Omega}{c}\alpha
  \left\{
    \left[
      \mathsf{N}_{\mathbf{k}}^{\Tra}
    \right]_{l-1,m}
    +
    \left[
      \mathsf{N}_{\mathbf{k}}^{\Tra}
    \right]_{l+1,m}
  \right\}.
\end{align}

Substituting those result into the boundary matching equation \eqref{eq:boundary_conditions_p_pol},
we can reach a matrix equation,
\begin{align}
  \mathsf{M}_{\mathbf{k}}^{\Inc}
  \mathbb{H}_{p,\mathbf{k}}^{\mathrm{inc}}
  +
  \mathsf{M}_{\mathbf{k}}^{\Refl}
  \mathbb{H}_{p,\mathbf{k}}^{\mathrm{ref}}
  -
  (\mathsf{M}_{\mathbf{k}}^{\Tra}
  +
  \widetilde{\mathsf{L}}_{\mathbf{k}})
  \mathbb{H}_{p,\mathbf{k}}^{\mathrm{tra}}
  &= 0.
\end{align}

Finally,
we can obtain the matrix equation which determine the reflection and transmission matrices,
\begin{align}
  \begin{pmatrix}
    \mathsf{M}_{\mathbf{k}}^{\Refl} 
    &
    -\mathsf{M}_{\mathbf{k}}^{\Tra}
    \\
    \widetilde{\mathsf{N}}_{\mathbf{k}}^{\Refl} 
    &
    -\widetilde{\mathsf{N}}_{\mathbf{k}}^{\Tra} 
  \end{pmatrix}
  \begin{pmatrix}
    \mathbb{H}_{p,\mathbf{k}}^{\Refl}
    \\
    \mathbb{H}_{p,\mathbf{k}}^{\Tra}
  \end{pmatrix}
  = 
  \begin{pmatrix}
    -\mathsf{M}_{\mathbf{k}}^{\Inc} 
    \mathbb{H}_{p,\mathbf{k}}^{\Inc}
    \\
    -\widetilde{\mathsf{N}}_{\mathbf{k}}^{\Inc} 
    \mathbb{H}_{p,\mathbf{k}}{\Inc}
  \end{pmatrix}.
  \tag{\ref{eq:p_pol}}
\end{align}
It is worth reminding that the characteristic impedance matrix,
\begin{align}
  \mathsf{Z}_{p,\mathbf{k}}^\tau 
  &=
  \operatorname{diag}
  (
  \cdots,
  Z_{p,\mathbf{k}_{-1}}^\tau,
  Z_{p,\mathbf{k}_{0}}^\tau,
  Z_{p,\mathbf{k}_{+1}}^\tau,
  \cdots
  ),
\end{align}
enables us to convert the reflection matrix calculated by means of the magnetic field, 
\begin{align}
  \mathbb{H}_{p,\mathbf{k}}^{\Refl}
  &=
  \mathsf{R}_{p,\mathbf{k}}
  \mathbb{H}_{p,\mathbf{k}}^{\Inc},
  \\
  \mathbb{E}_{p,\mathbf{k}}^{\Refl}
  &=
  \mathsf{Z}_{p,\mathbf{k}}^\ssg
  \mathsf{R}_{p,\mathbf{k}}
  \mathsf{Z}_{p,\mathbf{k}}^{\ssg\inv}
  \mathbb{E}_{p,\mathbf{k}}^{\Inc},
\end{align}
and the transmission matrix,
\begin{align}
  \mathbb{H}_{p,\mathbf{k}}^{\Tra}
  &=
  \mathsf{T}_{p,\mathbf{k}}
  \mathbb{H}_{p,\mathbf{k}}^{\Inc},
  \\
  \mathbb{E}_{p,\mathbf{k}}^{\Tra}
  &=
  \mathsf{Z}_{p,\mathbf{k}}^\ssl
  \mathsf{T}_{p,\mathbf{k}}
  \mathsf{Z}_{p,\mathbf{k}}^{\ssg\inv}
  \mathbb{E}_{p,\mathbf{k}}^{\Inc}.
\end{align}

\section{Recover the Fresnel equations}
\label{app:Fresnel}
In the flat boundary limit ($A \rightarrow 0$),
all the Bessel functions except the 0th order vanish,
\begin{align}
  \lim_{A\rightarrow 0}
  J_{l-m}(\phi_{\mathbf{k}_m}^{\sigma\tau}) 
  &=
  \delta_{l,m},
\end{align}
and the matrices become diagonal
\begin{align}
  \begin{cases}{}
  [\mathsf{M}_{\mathbf{k}}^{\sigma\tau}]_{lm}
  &\longrightarrow
  \cfrac{k_{x,m}}{k_{\parallel,m}}
  \operatorname{sgn}(\omega_m)
  \delta_{l,m}
  \\
  [\mathsf{N}_{\mathbf{k}}^{\sigma\tau}]_{lm}
  &\longrightarrow
  \cfrac{k_{x,m}}{k_{\parallel,m}}
  \cfrac{
    \sigma K_{\mathbf{k}_m}^\tau
  }{|k_{0,m}|}
  \delta_{l,m}
  \quad
  \end{cases}
  \quad
  (A\longrightarrow0).
\end{align}
Therefore,
the matrix equations (\ref{eq:RTs}, \ref{eq:RTp}) are reduced and give the conventional Fresnel coefficients,
  \begin{align}
    \begin{pmatrix}
      R_{s,\mathbf{k}}
      \\
      T_{s,\mathbf{k}}
    \end{pmatrix}
  &\longrightarrow
  \begin{pmatrix}
    \cfrac{k_x}{k_\parallel}
    &
    -\cfrac{k_x}{k_\parallel}
    \\
    \cfrac{
      k_x
      K_{\mathbf{k}}^\ssg
      }{
      k_\parallel
      k_0
    }
  &
  \cfrac{
    k_x
    K_{\mathbf{k}}^\ssl
    }{
    k_\parallel
    k_0
  }
\end{pmatrix}^{-1}
\begin{pmatrix}
  -\cfrac{k_x}{k_\parallel}
  \\
  \cfrac{
    k_x
    K_{\mathbf{k}}^\ssg
    }{
    k_\parallel
    k_0
  }
\end{pmatrix}
\\
&=
\begin{pmatrix}
  \cfrac{
    K_{\mathbf{k}}^\ssg
    -
    K_{\mathbf{k}}^\ssl
    }{
    K_{\mathbf{k}}^\ssg
    +
    K_{\mathbf{k}}^\ssl
  }
  \\
  \cfrac{
    2K_{\mathbf{k}}^\ssg
    }{
    K_{\mathbf{k}}^\ssg
    +
    K_{\mathbf{k}}^\ssl
  }
\end{pmatrix},
\\
\begin{pmatrix}
  R_{p,\mathbf{k}}
  \\
  T_{p,\mathbf{k}}
\end{pmatrix}
  &\longrightarrow
  \begin{pmatrix}
    \cfrac{k_x}{k_\parallel}
    &
    -\cfrac{k_x}{k_\parallel}
    \\
    \cfrac{
      k_x
      K_{\mathbf{k}}^\ssg
      }{
      k_\parallel
      k_0\epsilon^\ssg
    }
  &
  \cfrac{
    k_x
    K_{\mathbf{k}}^\ssl
    }{
    k_\parallel
    k_0\epsilon^\ssl
  }
\end{pmatrix}^{-1}
\begin{pmatrix}
  -\cfrac{k_x}{k_\parallel}
  \\
  \cfrac{
    k_x
    K_{\mathbf{k}}^\ssg
    }{
    k_\parallel
    k_0\epsilon^\ssg
  }
\end{pmatrix}
\\
&=
\begin{pmatrix}
  \cfrac{
    K_{\mathbf{k}}^\ssg/\epsilon^\ssg
    -
    K_{\mathbf{k}}^\ssl/\epsilon^\ssl
    }{
    K_{\mathbf{k}}^\ssg/\epsilon^\ssg
    +
    K_{\mathbf{k}}^\ssl/\epsilon^\ssl
  }
  \\
  \cfrac{
    2K_{\mathbf{k}}^\ssg/\epsilon^\ssg
    }{
    K_{\mathbf{k}}^\ssg/\epsilon^\ssg
    +
    K_{\mathbf{k}}^\ssl/\epsilon^\ssl
  }
\end{pmatrix}.
\end{align}


\begin{thebibliography}{32}%
\makeatletter
\providecommand \@ifxundefined [1]{%
 \@ifx{#1\undefined}
}%
\providecommand \@ifnum [1]{%
 \ifnum #1\expandafter \@firstoftwo
 \else \expandafter \@secondoftwo
 \fi
}%
\providecommand \@ifx [1]{%
 \ifx #1\expandafter \@firstoftwo
 \else \expandafter \@secondoftwo
 \fi
}%
\providecommand \natexlab [1]{#1}%
\providecommand \enquote  [1]{``#1''}%
\providecommand \bibnamefont  [1]{#1}%
\providecommand \bibfnamefont [1]{#1}%
\providecommand \citenamefont [1]{#1}%
\providecommand \href@noop [0]{\@secondoftwo}%
\providecommand \href [0]{\begingroup \@sanitize@url \@href}%
\providecommand \@href[1]{\@@startlink{#1}\@@href}%
\providecommand \@@href[1]{\endgroup#1\@@endlink}%
\providecommand \@sanitize@url [0]{\catcode `\\12\catcode `\$12\catcode
  `\&12\catcode `\#12\catcode `\^12\catcode `\_12\catcode `\%12\relax}%
\providecommand \@@startlink[1]{}%
\providecommand \@@endlink[0]{}%
\providecommand \url  [0]{\begingroup\@sanitize@url \@url }%
\providecommand \@url [1]{\endgroup\@href {#1}{\urlprefix }}%
\providecommand \urlprefix  [0]{URL }%
\providecommand \Eprint [0]{\href }%
\providecommand \doibase [0]{https://doi.org/}%
\providecommand \selectlanguage [0]{\@gobble}%
\providecommand \bibinfo  [0]{\@secondoftwo}%
\providecommand \bibfield  [0]{\@secondoftwo}%
\providecommand \translation [1]{[#1]}%
\providecommand \BibitemOpen [0]{}%
\providecommand \bibitemStop [0]{}%
\providecommand \bibitemNoStop [0]{.\EOS\space}%
\providecommand \EOS [0]{\spacefactor3000\relax}%
\providecommand \BibitemShut  [1]{\csname bibitem#1\endcsname}%
\let\auto@bib@innerbib\@empty
\bibitem [{\citenamefont {Sounas}\ and\ \citenamefont
  {Al{\`u}}(2017)}]{sounas2017non}%
  \BibitemOpen
  \bibfield  {author} {\bibinfo {author} {\bibfnamefont {D.~L.}\ \bibnamefont
  {Sounas}}\ and\ \bibinfo {author} {\bibfnamefont {A.}~\bibnamefont
  {Al{\`u}}},\ }\bibfield  {title} {\bibinfo {title} {Non-reciprocal photonics
  based on time modulation},\ }\href@noop {} {\bibfield  {journal} {\bibinfo
  {journal} {Nature Photonics}\ }\textbf {\bibinfo {volume} {11}},\ \bibinfo
  {pages} {774} (\bibinfo {year} {2017})}\BibitemShut {NoStop}%
\bibitem [{\citenamefont {Caloz}\ \emph {et~al.}(2018)\citenamefont {Caloz},
  \citenamefont {Al{\`u}}, \citenamefont {Tretyakov}, \citenamefont {Sounas},
  \citenamefont {Achouri},\ and\ \citenamefont
  {Deck-L{\'e}ger}}]{caloz2018electromagnetic}%
  \BibitemOpen
  \bibfield  {author} {\bibinfo {author} {\bibfnamefont {C.}~\bibnamefont
  {Caloz}}, \bibinfo {author} {\bibfnamefont {A.}~\bibnamefont {Al{\`u}}},
  \bibinfo {author} {\bibfnamefont {S.}~\bibnamefont {Tretyakov}}, \bibinfo
  {author} {\bibfnamefont {D.}~\bibnamefont {Sounas}}, \bibinfo {author}
  {\bibfnamefont {K.}~\bibnamefont {Achouri}},\ and\ \bibinfo {author}
  {\bibfnamefont {Z.-L.}\ \bibnamefont {Deck-L{\'e}ger}},\ }\bibfield  {title}
  {\bibinfo {title} {Electromagnetic nonreciprocity},\ }\href@noop {}
  {\bibfield  {journal} {\bibinfo  {journal} {Physical Review Applied}\
  }\textbf {\bibinfo {volume} {10}},\ \bibinfo {pages} {047001} (\bibinfo
  {year} {2018})}\BibitemShut {NoStop}%
\bibitem [{\citenamefont {Shaltout}\ \emph {et~al.}(2019)\citenamefont
  {Shaltout}, \citenamefont {Shalaev},\ and\ \citenamefont
  {Brongersma}}]{shaltout2019spatiotemporal}%
  \BibitemOpen
  \bibfield  {author} {\bibinfo {author} {\bibfnamefont {A.~M.}\ \bibnamefont
  {Shaltout}}, \bibinfo {author} {\bibfnamefont {V.~M.}\ \bibnamefont
  {Shalaev}},\ and\ \bibinfo {author} {\bibfnamefont {M.~L.}\ \bibnamefont
  {Brongersma}},\ }\bibfield  {title} {\bibinfo {title} {Spatiotemporal light
  control with active metasurfaces},\ }\href@noop {} {\bibfield  {journal}
  {\bibinfo  {journal} {Science}\ }\textbf {\bibinfo {volume} {364}} (\bibinfo
  {year} {2019})}\BibitemShut {NoStop}%
\bibitem [{\citenamefont {Galiffi}\ \emph {et~al.}(2019)\citenamefont
  {Galiffi}, \citenamefont {Huidobro},\ and\ \citenamefont
  {Pendry}}]{galiffi2019broadband}%
  \BibitemOpen
  \bibfield  {author} {\bibinfo {author} {\bibfnamefont {E.}~\bibnamefont
  {Galiffi}}, \bibinfo {author} {\bibfnamefont {P.}~\bibnamefont {Huidobro}},\
  and\ \bibinfo {author} {\bibfnamefont {J.}~\bibnamefont {Pendry}},\
  }\bibfield  {title} {\bibinfo {title} {Broadband nonreciprocal amplification
  in luminal metamaterials},\ }\href@noop {} {\bibfield  {journal} {\bibinfo
  {journal} {Physical Review Letters}\ }\textbf {\bibinfo {volume} {123}},\
  \bibinfo {pages} {206101} (\bibinfo {year} {2019})}\BibitemShut {NoStop}%
\bibitem [{\citenamefont {Huidobro}\ \emph {et~al.}(2019)\citenamefont
  {Huidobro}, \citenamefont {Galiffi}, \citenamefont {Guenneau}, \citenamefont
  {Craster},\ and\ \citenamefont {Pendry}}]{huidobro2019fresnel}%
  \BibitemOpen
  \bibfield  {author} {\bibinfo {author} {\bibfnamefont {P.~A.}\ \bibnamefont
  {Huidobro}}, \bibinfo {author} {\bibfnamefont {E.}~\bibnamefont {Galiffi}},
  \bibinfo {author} {\bibfnamefont {S.}~\bibnamefont {Guenneau}}, \bibinfo
  {author} {\bibfnamefont {R.~V.}\ \bibnamefont {Craster}},\ and\ \bibinfo
  {author} {\bibfnamefont {J.}~\bibnamefont {Pendry}},\ }\bibfield  {title}
  {\bibinfo {title} {{F}resnel drag in space--time-modulated metamaterials},\
  }\href@noop {} {\bibfield  {journal} {\bibinfo  {journal} {Proceedings of the
  National Academy of Sciences}\ }\textbf {\bibinfo {volume} {116}},\ \bibinfo
  {pages} {24943} (\bibinfo {year} {2019})}\BibitemShut {NoStop}%
\bibitem [{\citenamefont {Goedecke}\ and\ \citenamefont
  {O'Brien}(1988)}]{goedecke1988scattering}%
  \BibitemOpen
  \bibfield  {author} {\bibinfo {author} {\bibfnamefont {G.~H.}\ \bibnamefont
  {Goedecke}}\ and\ \bibinfo {author} {\bibfnamefont {S.~G.}\ \bibnamefont
  {O'Brien}},\ }\bibfield  {title} {\bibinfo {title} {Scattering by irregular
  inhomogeneous particles via the digitized green's function algorithm},\
  }\href@noop {} {\bibfield  {journal} {\bibinfo  {journal} {Applied Optics}\
  }\textbf {\bibinfo {volume} {27}},\ \bibinfo {pages} {2431} (\bibinfo {year}
  {1988})}\BibitemShut {NoStop}%
\bibitem [{\citenamefont {Yurkin}\ and\ \citenamefont
  {Hoekstra}(2007)}]{Yurkin_2007}%
  \BibitemOpen
  \bibfield  {author} {\bibinfo {author} {\bibfnamefont {M.}~\bibnamefont
  {Yurkin}}\ and\ \bibinfo {author} {\bibfnamefont {A.}~\bibnamefont
  {Hoekstra}},\ }\bibfield  {title} {\bibinfo {title} {The discrete dipole
  approximation: An overview and recent developments},\ }\href@noop {}
  {\bibfield  {journal} {\bibinfo  {journal} {Journal of Quantitative
  Spectroscopy and Radiative Transfer}\ }\textbf {\bibinfo {volume} {106}},\
  \bibinfo {pages} {558} (\bibinfo {year} {2007})}\BibitemShut {NoStop}%
\bibitem [{\citenamefont {Mie}(1908)}]{mie1908beitrage}%
  \BibitemOpen
  \bibfield  {author} {\bibinfo {author} {\bibfnamefont {G.}~\bibnamefont
  {Mie}},\ }\bibfield  {title} {\bibinfo {title} {Beitr{\"a}ge zur optik
  tr{\"u}ber medien, speziell kolloidaler metall{\"o}sungen},\ }\href@noop {}
  {\bibfield  {journal} {\bibinfo  {journal} {Annalen der physik}\ }\textbf
  {\bibinfo {volume} {330}},\ \bibinfo {pages} {377} (\bibinfo {year}
  {1908})}\BibitemShut {NoStop}%
\bibitem [{\citenamefont {Chandezon}\ \emph {et~al.}(1980)\citenamefont
  {Chandezon}, \citenamefont {Raoult},\ and\ \citenamefont
  {Maystre}}]{chandezon1980new}%
  \BibitemOpen
  \bibfield  {author} {\bibinfo {author} {\bibfnamefont {J.}~\bibnamefont
  {Chandezon}}, \bibinfo {author} {\bibfnamefont {G.}~\bibnamefont {Raoult}},\
  and\ \bibinfo {author} {\bibfnamefont {D.}~\bibnamefont {Maystre}},\
  }\bibfield  {title} {\bibinfo {title} {A new theoretical method for
  diffraction gratings and its numerical application},\ }\href@noop {}
  {\bibfield  {journal} {\bibinfo  {journal} {Journal of Optics}\ }\textbf
  {\bibinfo {volume} {11}},\ \bibinfo {pages} {235} (\bibinfo {year}
  {1980})}\BibitemShut {NoStop}%
\bibitem [{\citenamefont {Chandezon}\ \emph {et~al.}(1982)\citenamefont
  {Chandezon}, \citenamefont {Dupuis}, \citenamefont {Cornet},\ and\
  \citenamefont {Maystre}}]{chandezon1982multicoated}%
  \BibitemOpen
  \bibfield  {author} {\bibinfo {author} {\bibfnamefont {J.}~\bibnamefont
  {Chandezon}}, \bibinfo {author} {\bibfnamefont {M.}~\bibnamefont {Dupuis}},
  \bibinfo {author} {\bibfnamefont {G.}~\bibnamefont {Cornet}},\ and\ \bibinfo
  {author} {\bibfnamefont {D.}~\bibnamefont {Maystre}},\ }\bibfield  {title}
  {\bibinfo {title} {Multicoated gratings: a differential formalism applicable
  in the entire optical region},\ }\href@noop {} {\bibfield  {journal}
  {\bibinfo  {journal} {JOSA}\ }\textbf {\bibinfo {volume} {72}},\ \bibinfo
  {pages} {839} (\bibinfo {year} {1982})}\BibitemShut {NoStop}%
\bibitem [{\citenamefont {Li}(1994)}]{li1994multilayer}%
  \BibitemOpen
  \bibfield  {author} {\bibinfo {author} {\bibfnamefont {L.}~\bibnamefont
  {Li}},\ }\bibfield  {title} {\bibinfo {title} {Multilayer-coated diffraction
  gratings: differential method of chandezon et al. revisited},\ }\href@noop {}
  {\bibfield  {journal} {\bibinfo  {journal} {JOSA A}\ }\textbf {\bibinfo
  {volume} {11}},\ \bibinfo {pages} {2816} (\bibinfo {year}
  {1994})}\BibitemShut {NoStop}%
\bibitem [{\citenamefont {Li}\ and\ \citenamefont
  {Chandezon}(1996)}]{li1996improvement}%
  \BibitemOpen
  \bibfield  {author} {\bibinfo {author} {\bibfnamefont {L.}~\bibnamefont
  {Li}}\ and\ \bibinfo {author} {\bibfnamefont {J.}~\bibnamefont {Chandezon}},\
  }\bibfield  {title} {\bibinfo {title} {Improvement of the coordinate
  transformation method for surface-relief gratings with sharp edges},\
  }\href@noop {} {\bibfield  {journal} {\bibinfo  {journal} {JOSA A}\ }\textbf
  {\bibinfo {volume} {13}},\ \bibinfo {pages} {2247} (\bibinfo {year}
  {1996})}\BibitemShut {NoStop}%
\bibitem [{\citenamefont {Li}(1996)}]{li1996use}%
  \BibitemOpen
  \bibfield  {author} {\bibinfo {author} {\bibfnamefont {L.}~\bibnamefont
  {Li}},\ }\bibfield  {title} {\bibinfo {title} {Use of fourier series in the
  analysis of discontinuous periodic structures},\ }\href@noop {} {\bibfield
  {journal} {\bibinfo  {journal} {JOSA A}\ }\textbf {\bibinfo {volume} {13}},\
  \bibinfo {pages} {1870} (\bibinfo {year} {1996})}\BibitemShut {NoStop}%
\bibitem [{\citenamefont {Barnes}\ \emph {et~al.}(1995)\citenamefont {Barnes},
  \citenamefont {Preist}, \citenamefont {Kitson}, \citenamefont {Sambles},
  \citenamefont {Cotter},\ and\ \citenamefont {Nash}}]{barnes1995photonic}%
  \BibitemOpen
  \bibfield  {author} {\bibinfo {author} {\bibfnamefont {W.}~\bibnamefont
  {Barnes}}, \bibinfo {author} {\bibfnamefont {T.}~\bibnamefont {Preist}},
  \bibinfo {author} {\bibfnamefont {S.}~\bibnamefont {Kitson}}, \bibinfo
  {author} {\bibfnamefont {J.}~\bibnamefont {Sambles}}, \bibinfo {author}
  {\bibfnamefont {N.}~\bibnamefont {Cotter}},\ and\ \bibinfo {author}
  {\bibfnamefont {D.}~\bibnamefont {Nash}},\ }\bibfield  {title} {\bibinfo
  {title} {Photonic gaps in the dispersion of surface plasmons on gratings},\
  }\href@noop {} {\bibfield  {journal} {\bibinfo  {journal} {Physical Review
  B}\ }\textbf {\bibinfo {volume} {51}},\ \bibinfo {pages} {11164} (\bibinfo
  {year} {1995})}\BibitemShut {NoStop}%
\bibitem [{\citenamefont {Harris}\ \emph {et~al.}(1995)\citenamefont {Harris},
  \citenamefont {Preist},\ and\ \citenamefont
  {Sambles}}]{harris1995differential}%
  \BibitemOpen
  \bibfield  {author} {\bibinfo {author} {\bibfnamefont {J.}~\bibnamefont
  {Harris}}, \bibinfo {author} {\bibfnamefont {T.}~\bibnamefont {Preist}},\
  and\ \bibinfo {author} {\bibfnamefont {J.}~\bibnamefont {Sambles}},\
  }\bibfield  {title} {\bibinfo {title} {Differential formalism for multilayer
  diffraction gratings made with uniaxial materials},\ }\href@noop {}
  {\bibfield  {journal} {\bibinfo  {journal} {JOSA A}\ }\textbf {\bibinfo
  {volume} {12}},\ \bibinfo {pages} {1965} (\bibinfo {year}
  {1995})}\BibitemShut {NoStop}%
\bibitem [{\citenamefont {Barnes}\ \emph {et~al.}(1996)\citenamefont {Barnes},
  \citenamefont {Preist}, \citenamefont {Kitson},\ and\ \citenamefont
  {Sambles}}]{barnes1996physical}%
  \BibitemOpen
  \bibfield  {author} {\bibinfo {author} {\bibfnamefont {W.~L.}\ \bibnamefont
  {Barnes}}, \bibinfo {author} {\bibfnamefont {T.}~\bibnamefont {Preist}},
  \bibinfo {author} {\bibfnamefont {S.}~\bibnamefont {Kitson}},\ and\ \bibinfo
  {author} {\bibfnamefont {J.}~\bibnamefont {Sambles}},\ }\bibfield  {title}
  {\bibinfo {title} {Physical origin of photonic energy gaps in the propagation
  of surface plasmons on gratings},\ }\href@noop {} {\bibfield  {journal}
  {\bibinfo  {journal} {Physical Review B}\ }\textbf {\bibinfo {volume} {54}},\
  \bibinfo {pages} {6227} (\bibinfo {year} {1996})}\BibitemShut {NoStop}%
\bibitem [{\citenamefont {Harris}\ \emph {et~al.}(1996)\citenamefont {Harris},
  \citenamefont {Preist}, \citenamefont {Wood},\ and\ \citenamefont
  {Sambles}}]{harris1996conical}%
  \BibitemOpen
  \bibfield  {author} {\bibinfo {author} {\bibfnamefont {J.}~\bibnamefont
  {Harris}}, \bibinfo {author} {\bibfnamefont {T.}~\bibnamefont {Preist}},
  \bibinfo {author} {\bibfnamefont {E.}~\bibnamefont {Wood}},\ and\ \bibinfo
  {author} {\bibfnamefont {J.}~\bibnamefont {Sambles}},\ }\bibfield  {title}
  {\bibinfo {title} {Conical diffraction from multicoated gratings containing
  uniaxial materials},\ }\href@noop {} {\bibfield  {journal} {\bibinfo
  {journal} {JOSA A}\ }\textbf {\bibinfo {volume} {13}},\ \bibinfo {pages}
  {803} (\bibinfo {year} {1996})}\BibitemShut {NoStop}%
\bibitem [{\citenamefont {Kitamura}\ and\ \citenamefont
  {Murakami}(2013)}]{kitamura2013hermitian}%
  \BibitemOpen
  \bibfield  {author} {\bibinfo {author} {\bibfnamefont {Y.}~\bibnamefont
  {Kitamura}}\ and\ \bibinfo {author} {\bibfnamefont {S.}~\bibnamefont
  {Murakami}},\ }\bibfield  {title} {\bibinfo {title} {Hermitian two-band model
  for one-dimensional plasmonic crystals},\ }\href@noop {} {\bibfield
  {journal} {\bibinfo  {journal} {Physical Review B}\ }\textbf {\bibinfo
  {volume} {88}},\ \bibinfo {pages} {045406} (\bibinfo {year}
  {2013})}\BibitemShut {NoStop}%
\bibitem [{\citenamefont {Murtaza}\ \emph {et~al.}(2017)\citenamefont
  {Murtaza}, \citenamefont {Syed},\ and\ \citenamefont
  {Naqvi}}]{murtaza2017study}%
  \BibitemOpen
  \bibfield  {author} {\bibinfo {author} {\bibfnamefont {G.}~\bibnamefont
  {Murtaza}}, \bibinfo {author} {\bibfnamefont {A.~A.}\ \bibnamefont {Syed}},\
  and\ \bibinfo {author} {\bibfnamefont {Q.~A.}\ \bibnamefont {Naqvi}},\
  }\bibfield  {title} {\bibinfo {title} {Study of scattering from a periodic
  grating structure using lorentz--drude model and chandezon method},\
  }\href@noop {} {\bibfield  {journal} {\bibinfo  {journal} {Optik}\ }\textbf
  {\bibinfo {volume} {133}},\ \bibinfo {pages} {9} (\bibinfo {year}
  {2017})}\BibitemShut {NoStop}%
\bibitem [{\citenamefont {Ward}\ and\ \citenamefont
  {Pendry}(1996)}]{ward1996refraction}%
  \BibitemOpen
  \bibfield  {author} {\bibinfo {author} {\bibfnamefont {A.}~\bibnamefont
  {Ward}}\ and\ \bibinfo {author} {\bibfnamefont {J.~B.}\ \bibnamefont
  {Pendry}},\ }\bibfield  {title} {\bibinfo {title} {Refraction and geometry in
  maxwell's equations},\ }\href@noop {} {\bibfield  {journal} {\bibinfo
  {journal} {Journal of modern optics}\ }\textbf {\bibinfo {volume} {43}},\
  \bibinfo {pages} {773} (\bibinfo {year} {1996})}\BibitemShut {NoStop}%
\bibitem [{\citenamefont {Leonhardt}(2006)}]{leonhardt2006optical}%
  \BibitemOpen
  \bibfield  {author} {\bibinfo {author} {\bibfnamefont {U.}~\bibnamefont
  {Leonhardt}},\ }\bibfield  {title} {\bibinfo {title} {Optical conformal
  mapping},\ }\href@noop {} {\bibfield  {journal} {\bibinfo  {journal}
  {Science}\ }\textbf {\bibinfo {volume} {312}},\ \bibinfo {pages} {1777}
  (\bibinfo {year} {2006})}\BibitemShut {NoStop}%
\bibitem [{\citenamefont {Liu}\ \emph {et~al.}(2010)\citenamefont {Liu},
  \citenamefont {Zentgraf}, \citenamefont {Bartal},\ and\ \citenamefont
  {Zhang}}]{liu2010transformational}%
  \BibitemOpen
  \bibfield  {author} {\bibinfo {author} {\bibfnamefont {Y.}~\bibnamefont
  {Liu}}, \bibinfo {author} {\bibfnamefont {T.}~\bibnamefont {Zentgraf}},
  \bibinfo {author} {\bibfnamefont {G.}~\bibnamefont {Bartal}},\ and\ \bibinfo
  {author} {\bibfnamefont {X.}~\bibnamefont {Zhang}},\ }\bibfield  {title}
  {\bibinfo {title} {Transformational plasmon optics},\ }\href@noop {}
  {\bibfield  {journal} {\bibinfo  {journal} {Nano Letters}\ }\textbf {\bibinfo
  {volume} {10}},\ \bibinfo {pages} {1991} (\bibinfo {year}
  {2010})}\BibitemShut {NoStop}%
\bibitem [{\citenamefont {Vakil}\ and\ \citenamefont
  {Engheta}(2011)}]{vakil2011transformation}%
  \BibitemOpen
  \bibfield  {author} {\bibinfo {author} {\bibfnamefont {A.}~\bibnamefont
  {Vakil}}\ and\ \bibinfo {author} {\bibfnamefont {N.}~\bibnamefont
  {Engheta}},\ }\bibfield  {title} {\bibinfo {title} {Transformation optics
  using graphene},\ }\href@noop {} {\bibfield  {journal} {\bibinfo  {journal}
  {Science}\ }\textbf {\bibinfo {volume} {332}},\ \bibinfo {pages} {1291}
  (\bibinfo {year} {2011})}\BibitemShut {NoStop}%
\bibitem [{\citenamefont {Xu}\ and\ \citenamefont
  {Chen}(2015)}]{xu2015conformal}%
  \BibitemOpen
  \bibfield  {author} {\bibinfo {author} {\bibfnamefont {L.}~\bibnamefont
  {Xu}}\ and\ \bibinfo {author} {\bibfnamefont {H.}~\bibnamefont {Chen}},\
  }\bibfield  {title} {\bibinfo {title} {Conformal transformation optics},\
  }\href@noop {} {\bibfield  {journal} {\bibinfo  {journal} {Nature Photonics}\
  }\textbf {\bibinfo {volume} {9}},\ \bibinfo {pages} {15} (\bibinfo {year}
  {2015})}\BibitemShut {NoStop}%
\bibitem [{\citenamefont {Pendry}\ \emph {et~al.}(2015)\citenamefont {Pendry},
  \citenamefont {Luo},\ and\ \citenamefont {Zhao}}]{pendry2015transforming}%
  \BibitemOpen
  \bibfield  {author} {\bibinfo {author} {\bibfnamefont {J.}~\bibnamefont
  {Pendry}}, \bibinfo {author} {\bibfnamefont {Y.}~\bibnamefont {Luo}},\ and\
  \bibinfo {author} {\bibfnamefont {R.}~\bibnamefont {Zhao}},\ }\bibfield
  {title} {\bibinfo {title} {Transforming the optical landscape},\ }\href@noop
  {} {\bibfield  {journal} {\bibinfo  {journal} {Science}\ }\textbf {\bibinfo
  {volume} {348}},\ \bibinfo {pages} {521} (\bibinfo {year}
  {2015})}\BibitemShut {NoStop}%
\bibitem [{\citenamefont {Pendry}\ \emph {et~al.}(2019)\citenamefont {Pendry},
  \citenamefont {Huidobro},\ and\ \citenamefont {Ding}}]{pendry2019computing}%
  \BibitemOpen
  \bibfield  {author} {\bibinfo {author} {\bibfnamefont {J.}~\bibnamefont
  {Pendry}}, \bibinfo {author} {\bibfnamefont {P.~A.}\ \bibnamefont
  {Huidobro}},\ and\ \bibinfo {author} {\bibfnamefont {K.}~\bibnamefont
  {Ding}},\ }\bibfield  {title} {\bibinfo {title} {Computing one-dimensional
  metasurfaces},\ }\href@noop {} {\bibfield  {journal} {\bibinfo  {journal}
  {Physical Review B}\ }\textbf {\bibinfo {volume} {99}},\ \bibinfo {pages}
  {085408} (\bibinfo {year} {2019})}\BibitemShut {NoStop}%
\bibitem [{\citenamefont {Pendry}(2008)}]{pendry2008time}%
  \BibitemOpen
  \bibfield  {author} {\bibinfo {author} {\bibfnamefont {J.}~\bibnamefont
  {Pendry}},\ }\bibfield  {title} {\bibinfo {title} {Time reversal and negative
  refraction},\ }\href@noop {} {\bibfield  {journal} {\bibinfo  {journal}
  {Science}\ }\textbf {\bibinfo {volume} {322}},\ \bibinfo {pages} {71}
  (\bibinfo {year} {2008})}\BibitemShut {NoStop}%
\bibitem [{\citenamefont {Cuyt}\ \emph {et~al.}(2008)\citenamefont {Cuyt},
  \citenamefont {Petersen}, \citenamefont {Verdonk}, \citenamefont
  {Waadeland},\ and\ \citenamefont {Jones}}]{cuyt2008handbook}%
  \BibitemOpen
  \bibfield  {author} {\bibinfo {author} {\bibfnamefont {A.~A.}\ \bibnamefont
  {Cuyt}}, \bibinfo {author} {\bibfnamefont {V.}~\bibnamefont {Petersen}},
  \bibinfo {author} {\bibfnamefont {B.}~\bibnamefont {Verdonk}}, \bibinfo
  {author} {\bibfnamefont {H.}~\bibnamefont {Waadeland}},\ and\ \bibinfo
  {author} {\bibfnamefont {W.~B.}\ \bibnamefont {Jones}},\ }\href@noop {}
  {\emph {\bibinfo {title} {Handbook of continued fractions for special
  functions}}}\ (\bibinfo  {publisher} {Springer Science \& Business Media},\
  \bibinfo {year} {2008})\BibitemShut {NoStop}%
\bibitem [{\citenamefont {Li}(1999)}]{li1999justification}%
  \BibitemOpen
  \bibfield  {author} {\bibinfo {author} {\bibfnamefont {L.}~\bibnamefont
  {Li}},\ }\bibfield  {title} {\bibinfo {title} {Justification of matrix
  truncation in the modal methods of diffraction gratings},\ }\href@noop {}
  {\bibfield  {journal} {\bibinfo  {journal} {Journal of Optics A: Pure and
  Applied Optics}\ }\textbf {\bibinfo {volume} {1}},\ \bibinfo {pages} {531}
  (\bibinfo {year} {1999})}\BibitemShut {NoStop}%
\bibitem [{\citenamefont {Shcherbakov}\ and\ \citenamefont
  {Tishchenko}(2013)}]{shcherbakov2013efficient}%
  \BibitemOpen
  \bibfield  {author} {\bibinfo {author} {\bibfnamefont {A.~A.}\ \bibnamefont
  {Shcherbakov}}\ and\ \bibinfo {author} {\bibfnamefont {A.~V.}\ \bibnamefont
  {Tishchenko}},\ }\bibfield  {title} {\bibinfo {title} {Efficient curvilinear
  coordinate method for grating diffraction simulation},\ }\href@noop {}
  {\bibfield  {journal} {\bibinfo  {journal} {Optics express}\ }\textbf
  {\bibinfo {volume} {21}},\ \bibinfo {pages} {25236} (\bibinfo {year}
  {2013})}\BibitemShut {NoStop}%
\bibitem [{Note1()}]{Note1}%
  \BibitemOpen
  \bibinfo {note} {See the Supplemental Materials.}\BibitemShut {Stop}%
\bibitem [{\citenamefont {Wood}(1902)}]{wood1902xlii}%
  \BibitemOpen
  \bibfield  {author} {\bibinfo {author} {\bibfnamefont {R.~W.}\ \bibnamefont
  {Wood}},\ }\bibfield  {title} {\bibinfo {title} {{XLII}. {O}n a remarkable
  case of uneven distribution of light in a diffraction grating spectrum},\
  }\href@noop {} {\bibfield  {journal} {\bibinfo  {journal} {The London,
  Edinburgh, and Dublin Philosophical Magazine and Journal of Science}\
  }\textbf {\bibinfo {volume} {4}},\ \bibinfo {pages} {396} (\bibinfo {year}
  {1902})}\BibitemShut {NoStop}%
\end{thebibliography}
%

\end{document}